\documentclass[12pt]{article}
\newcommand{\christ}{\left\{{\lambda \hfill}\atop{\mu\nu}\right\}}
\newcommand{\ft}[4]{F^{\hat{#1}\hat{#2}}{}_{\hat{#3}\hat{#4}}}
\begin{document}

\title{ Spherically-symmetric gravitational fields in the metric-affine 
gauge theory of gravitation}

\author{A. V. Minkevich$^{1,2}$, Yu. G. Vasilevski$^1$}
\date{}
\maketitle 

$^1$Department of Theoretical Physics, Belarussian State University, 
Minsk, Belarus.  

$^2$Department of Physics and Computer Methods, University of Warmia and 
Mazury in Olsztyn, Poland.

E-mail: minkav@bsu.by; awm@matman.uwm.edu.pl

\vspace{1.0cm}

\begin{abstract}
 Geometric structure of spherically-symmetric space-time in 
metric-affine gauge theory of gravity is studied.  
Restrictions on curvature tensor and Bianchi identities are obtained. By 
using certain simple gravitational Lagrangian the solution of gravitational 
equations for vacuum spherically-symmetric 
gravitational field is obtained.\\
\\
{\raggedright  gr-qc/0301098}
\end{abstract}

\vspace{1.0cm}

As it is known, the application of gauge approach to gravitational 
interaction leads to generalization of Einsteinian theory of gravitation. At 
present there are different gauge theories of gravitation in dependence on 
using gauge group corresponding to gravitational interaction. The 
metric-affine gauge theory of gravitation (MAGT) is one of the most general
gauge theories of gravity and it is based on the group of 
affine transformations A(4,R) as gauge groupe \cite{Hehl}. In MAGT 
spase-time continuum possesses curvature, torsion and nonmetricity, and as 
sources of gravitational field are energy-momrntum tensor and so-called 
hypermomentum which is generalization of spin-momentum tensor of Poincare 
gauge theory of gravitation (PGT).       
 
The system of gravitational equations of MAGT is  
complicated system of nonlinear differential equations. Their analysis is 
simplified in the case of models with high 
space symmetries. Homogeneous isotropic models in MAGT were investigated in 
Ref. \cite{Gar}. In the framework of MAGT spherically-symmetric models are 
analyzed in present paper.

The geometric structure of space-time in MAGT is determined by three 
tensors: metrics $g_{\mu\nu}$, torsion $S^{\lambda}{}_{\mu\nu}$ and 
nonmetricity $Q_{\lambda\mu\nu}$\footnote{$\mu$, $\nu$, ... are holonomic 
indices; i, k, ... are anholonomic (tetrad) indices. Numerical tetrad 
indices are denoted by means of a sign \^\  over them.}. 
By using the system of spherical 
coordinates ($x^{0}=ct$, $x^{1}=r$, $x^{2}=\theta$, $x^{3}=\varphi$), we 
write metrics in the following form
\begin{equation}
g_{\mu\nu} ={\rm diag}( e^{\nu} , -e^{\lambda} , -r^{2} 
, -r^{2}\sin^{2}\theta ), 
\end{equation}
where $\nu=\nu(r,t)$, $\lambda=\lambda(r,t)$ are two functions of radial 
coordinate r and time t. The structure of tensors 
$S^{\lambda}{}_{\mu\nu}$ and $Q_{\lambda\mu\nu}$ in spherically-symmetric 
case was studied in Ref. \cite{Mink}. The torsion 
is determined by 8 functions $S_{i}=S_{i}(r,t)$ ($i=1,\ 2,\ \ldots 8$) and 
nonmetricity --- by 12 functions $Q_{k}=Q_{k}(r,t)$ ($k = 0,\ 1,\ \ldots 
11$). Namely nonvanishing components of 
tensors $S_{\lambda\mu\nu}$ and $Q_{\lambda\mu\nu}$ are: 
\begin{equation}
\begin{array}{c}
\vspace{0.35 cm}
S_{001}=S_{1},\ S_{212}=S_{2},\ S_{101}=S_{3},\ S_{202}=S_{4},\
S_{313}=S_{2}\sin^{2}\theta,\\
\vspace{0.35 cm}
S_{303}=S_{4}\sin^{2}\theta,\
S_{032}=S_{5}\sin\theta,\ 
S_{132}=S_{6}\sin\theta,\\
\vspace{0.35 cm}
S_{302}=-S_{203}=S_{7}\sin\theta,\ 
S_{312}=-S_{213}=S_{8}\sin\theta,\ 
\end{array}
\end{equation}
\begin{equation}
\begin{array}{c}
\vspace{0.35 cm}
Q_{000}=Q_{0},\ Q_{001}=Q_{1},\ Q_{010}=Q_{2},\ Q_{011}=Q_{3},\
Q_{110}=Q_{4},\\
\vspace{0.35 cm}
Q_{111}=Q_{5},\ 
Q_{022}=Q_{6},\ Q_{122}=Q_{7},\
Q_{220}=Q_{8},\ Q_{221}=Q_{9},\\ 
\vspace{0.35 cm}
Q_{023}=-Q_{032}=Q_{10}\sin\theta,\
Q_{123}=-Q_{132}=Q_{11}\sin\theta,\\ 
\vspace{0.35 cm}
Q_{033}=Q_{6}\sin^{2}\theta,\
Q_{133}=Q_{7}\sin^{2}\theta,\
Q_{330}=Q_{8}\sin^{2}\theta,\ 
Q_{331}=Q_{9}\sin^{2}\theta.\\
\end{array}
\end{equation}
Note that functions $S_{i}$ ($i=5,\ 6,\ 7,\ 8$) and $Q_{k}$ ($k = 10,\ 11$)
have pseudoscalar character. All other components of tensors
 $S_{\lambda\mu\nu}$ and $Q_{\lambda\mu\nu}$ vanish, with the exception 
of components connected with components $(2)-(3)$ by symmetry properties of 
torsion
$S_{\lambda\mu\nu}=-S_{\lambda\nu\mu}$ and nonmetricity 
$Q_{\lambda\mu\nu}=Q_{\mu\lambda\nu}$.

 By choosing diagonal tetrad $h^{i}{}_{\mu}$ corresponding to metrics (1) 
\begin{equation}
h^{i}{}_{\mu} = {\rm diag}(e^{\frac{\nu}{2}},\  
e^{\frac{\lambda}{2}},\  r,\ 
 r\sin\theta),
\end{equation}
we find anholonomic connection:
\begin{equation}
A^{ik}{}_{\mu} =  h^{k\nu}(\partial_{\mu}h^{i}{}_{\nu}-
h^{i}{}_{\lambda}\Gamma^{\lambda}{}_{\nu\mu}),
\end{equation}
where holonomic connection $\Gamma^{\lambda}{}_{\mu\nu} = \christ + 
S^{\lambda}{}_{\mu\nu} + 
S_{\mu\nu}{}^{\lambda} + S_{\nu\mu}{}^{\lambda} + 
\frac{1}{2}(Q_{\mu\nu}{}^{\lambda} - Q_{\mu}{}^{\lambda}{}_{\nu} - 
Q_{\nu}{}^{\lambda}{}_{\mu})$ and $\christ$ are Christoffel symbols. 
Nonvanishing components of connection $A^{ik}{}_{\mu}$ are:

\vspace{0.35 cm}
\begin{displaymath}
\begin{array}{c}
\vspace{0.35 cm}
A^{\hat{0}\hat{0}}{}_{0}=A_{0},\
A^{\hat{0}\hat{0}}{}_{1}=A_{1},\
A^{\hat{1}\hat{0}}{}_{0}=A_{2},\
A^{\hat{1}\hat{0}}{}_{1}=A_{3},\
A^{\hat{2}\hat{0}}{}_{2}=A_{4},\\
\vspace{0.35 cm}
A^{\hat{3}\hat{0}}{}_{3}=A_{4}\sin\theta,\
A^{\hat{0}\hat{1}}{}_{0}=A_{5},\
A^{\hat{0}\hat{1}}{}_{1}=A_{6},\
A^{\hat{1}\hat{1}}{}_{0}=A_{7},\\
\vspace{0.35 cm}
A^{\hat{1}\hat{1}}{}_{1}=A_{8},\
A^{\hat{2}\hat{1}}{}_{2}=A_{9},\
A^{\hat{3}\hat{1}}{}_{3}=A_{9}\sin\theta,\
A^{\hat{0}\hat{2}}{}_{2}=A_{10},\\
\vspace{0.35 cm}
\end{array}
\end{displaymath}
\begin{equation}
\begin{array}{c}
\vspace{0.35 cm}
A^{\hat{0}\hat{3}}{}_{3}=A_{10}\sin\theta,\
A^{\hat{1}\hat{2}}{}_{2}=A_{11},\
A^{\hat{1}\hat{3}}{}_{3}=A_{11}\sin\theta,\
A^{\hat{2}\hat{2}}{}_{0}=A^{\hat{3}\hat{3}}{}_{0}=A_{12},\\
\vspace{0.35 cm}
A^{\hat{2}\hat{2}}{}_{1}=A^{\hat{3}\hat{3}}{}_{1}=A_{13},\
A^{\hat{0}\hat{3}}{}_{2}=A_{14},\
A^{\hat{0}\hat{2}}{}_{3}=-A_{14}\sin\theta,\
A^{\hat{1}\hat{3}}{}_{2}=A_{15},\\
\vspace{0.35 cm}
A^{\hat{1}\hat{2}}{}_{3}=-A_{15}\sin\theta,\
A^{\hat{2}\hat{3}}{}_{0}=-A^{\hat{3}\hat{2}}{}_{0}=A_{16},\
A^{\hat{2}\hat{3}}{}_{1}=-A^{\hat{3}\hat{2}}{}_{1}=A_{17},\\
\vspace{0.35 cm}
A^{\hat{3}\hat{0}}{}_{2}=A_{18},\
A^{\hat{2}\hat{0}}{}_{3}=-A_{18}\sin\theta,\
A^{\hat{3}\hat{1}}{}_{2}=A_{19},\
A^{\hat{2}\hat{1}}{}_{3}=-A_{19}\sin\theta,\\
\vspace{0.35 cm}
\end{array}
\end{equation}
where explisit form of functions $A_{i}$ ($i=0,\ 2,\ \ldots 19$) is:
\begin{equation}
\begin{array}{c}
\vspace{0.35 cm}
A_{0}=\frac{1}{2}e^{-\nu}Q_{0},\
A_{1}=\frac{1}{2}e^{-\nu}Q_{1},\
A_{2}=\frac{1}{2}e^{-\frac{1}{2}(\lambda+\nu)}(Q_{1}-
2Q_{2}+4S_{1}-e^{\nu}\nu'),\\
\vspace{0.35 cm}
A_{3}=-\frac{1}{2}e^{-\frac{1}{2}(\lambda+\nu)}(Q_{4}-
4S_{3}+e^{\lambda}\dot{\lambda}),\
A_{4}=-\frac{1}{2r}e^{-\frac{1}{2}\nu}(Q_{8}-4S_{4}),\\
\vspace{0.35 cm}
A_{5}=\frac{1}{2}e^{-\frac{1}{2}(\lambda+\nu)}(-Q_{1}-
4S_{1}+e^{\nu}\nu'),\
A_{6}=\frac{1}{2}e^{-\frac{1}{2}(\lambda+\nu)}(-2Q_{3}+
Q_{4}-\\
\vspace{0.35 cm}
4S_{3}+e^{\lambda}\dot{\lambda}),\
A_{7}=\frac{1}{2}e^{-\lambda}Q_{4},\
A_{8}=\frac{1}{2}e^{-\lambda}Q_{5},\
A_{9}=\frac{1}{2r}e^{-\frac{1}{2}\lambda}(2r+
Q_{9}-4S_{2}),\\
\vspace{0.35 cm}
A_{10}=\frac{1}{2r}e^{-\frac{1}{2}\nu}(Q_{8}-2Q_{6}-
4S_{4}),\\
\vspace{0.35 cm}
A_{11}=-\frac{1}{2r}e^{-\frac{1}{2}\lambda}(2r-2Q_{7}+
Q_{9}-4S_{2}),
A_{12}=\frac{1}{2r^{2}}Q_{8},\\
\vspace{0.35 cm}
A_{13}=\frac{1}{2r^{2}}Q_{9},\
A_{14}=\frac{1}{r}e^{-\frac{1}{2}\nu}S_{5},\
A_{15}=-\frac{1}{r}e^{-\frac{1}{2}\lambda}S_{6},\\
\vspace{0.35 cm}
A_{16}=\frac{1}{r^{2}}(Q_{10}-S_{5}-2S_{7}),\
A_{17}=\frac{1}{r^{2}}(Q_{11}-S_{6}-2S_{8}),\\
\vspace{0.35 cm}
A_{18}=\frac{1}{r}e^{-\frac{1}{2}\nu}(Q_{10}-S_{5}),\
A_{19}=-\frac{1}{r}e^{-\frac{1}{2}\lambda}(Q_{11}-S_{6}).\\
\vspace{0.35 cm}
\end{array}
\end{equation} 

 The curvature tensor can be calculated according to his definition:
\begin{equation}
F^{ik}{}_{\mu\nu}\ =\ 2\partial_{[\mu}A^{ik}{}_{\nu]}\ +\ 
2A^{i}{}_{l[\nu}A^{lk}{}_{\mu]}.
\end{equation}
In considered case the curvature is determined by 27 functions $F_{i}$ ($i 
= 0, 1, \ldots 26$) depending on functions $\nu$, $\lambda$, $S_{i}$, 
$Q_{k}$:
\begin{displaymath}
\begin{array}{c}
\vspace{0.35 cm}
\ft{0}{0}{1}{0}=F_{0},\ \ft{0}{1}{1}{0}=F_{1},\ 
\ft{0}{2}{2}{0}=\ft{0}{3}{3}{0}=F_{2},\
\ft{0}{2}{2}{1}=\ft{0}{3}{3}{1}=F_{3},\\
\vspace{0.35 cm}
\ft{1}{0}{1}{0}=F_{4},\
\ft{1}{1}{1}{0}=F_{5},\
\ft{1}{2}{2}{0}=\ft{1}{3}{3}{0}=F_{6},\
\ft{1}{2}{2}{1}=\ft{1}{3}{3}{1}=F_{7},\\
\vspace{0.35 cm} 
\ft{2}{0}{2}{0}=\ft{3}{0}{3}{0}=F_{8},\ 
\ft{2}{0}{2}{1}=\ft{3}{0}{3}{1}=F_{9},\
\ft{3}{1}{3}{0}=\ft{2}{1}{2}{0}=F_{10},\\
\vspace{0.35 cm}
\end{array}
\end{displaymath}
\begin{equation}
\begin{array}{c}
\vspace{0.35 cm}
\ft{3}{1}{3}{1}=\ft{2}{1}{2}{1}=F_{11},\
\ft{2}{2}{1}{0}=\ft{3}{3}{1}{0}=F_{12},\
-\ft{3}{2}{3}{2}=\ft{2}{3}{3}{2}=F_{13},\\
\vspace{0.35 cm}
\ft{0}{0}{3}{2}=F_{14},\
\ft{0}{1}{3}{2}=F_{15},\ \ft{1}{0}{3}{2}=F_{16},\ 
\ft{1}{1}{3}{2}=F_{17},\\
\vspace{0.35 cm}
\ft{2}{0}{3}{0}=-\ft{3}{0}{2}{0}=F_{18},\  
\ft{2}{0}{3}{1}=-\ft{3}{0}{2}{1}=F_{19},\ 
\ft{0}{2}{3}{0}=-\ft{0}{3}{2}{0}=F_{20},\\ 
\vspace{0.35 cm}
\ft{0}{2}{3}{1}=-\ft{0}{3}{2}{1}=F_{21},\ 
\ft{3}{1}{2}{0}=-\ft{2}{1}{3}{0}=F_{22},\
\ft{3}{1}{2}{1}=-\ft{2}{1}{3}{1}=F_{23},\\ 
\vspace{0.35 cm}
\ft{1}{2}{3}{1}=-\ft{1}{3}{2}{1}=F_{24},\  
\ft{1}{2}{3}{0}=-\ft{1}{3}{2}{0}=F_{25},\  
\ft{3}{2}{1}{0}=-\ft{2}{3}{1}{0}=F_{26},\\
\vspace{0.35 cm}
\ft{2}{2}{3}{2}=\ft{3}{3}{3}{2}=\frac{1}{2}(F_{14}-F_{17}).\\ 
\end{array}
\end{equation}
Explicit form of functions $F_{i}$ is

\begin{equation}
\begin{array}{c}
\vspace{0.35 cm}  
F_{0}=\frac{1}{2}e^{-\frac{3}{2}(\lambda+\nu)}[-4Q_{3}S_{1}-e^{\lambda}(
\dot{Q_{1}}-Q'_{0}+Q_{0}\nu'+Q_{2}\dot{\lambda}-\\
\vspace{0.35 cm}  
Q_{1}\dot{\nu})
+Q_{2}(2Q_{3}-Q_{4}+4S_{3})+Q_{3}(e^{\nu}\nu'-Q_{1})],\\
\vspace{0.35 cm}   
F_{1}=\frac{1}{4}e^{-(\lambda+2\nu)}[Q_{0}(Q_{4}-2Q_{3}-4S_{3}+e^{\lambda}
\dot{\lambda})+Q_{1}(Q_{1}+4S_{1}+e^{\nu}\lambda')]+\\
\vspace{0.35 cm}  
\frac{1}{4}e^{-(2\lambda+\nu)}
\{Q_{5}(Q_{1}+4S_{1})+Q^{2}_{4}+e^{\lambda}[4\dot{Q_{3}}-2\dot{Q_{4}}
+8\dot{S_{3}}-2Q_{1}'-8S_{1}'+\\
\vspace{0.35 cm}  
4S_{1}(\lambda'+\nu')-e^{\nu}(\lambda'\nu'
+\nu{'}^{2}+2\nu{'}{'})-(4S_{3}+2Q_{3}-Q_{4})(\dot{\lambda}+\dot{\nu})+\\
\vspace{0.35 cm}  
Q_{4}\dot{\lambda}-e^{\lambda}
(\dot{\lambda}^{2}+2\ddot{\lambda}-\dot{\lambda}\dot{\nu})]-Q_{4}(2Q_{3}+
4S_{3})-e^{\nu}Q_{5}\nu'\},\\
\vspace{0.35 cm}    
F_{2}=-\frac{1}{4r^{4}}e^{-(\lambda+\nu)}\{r^{2}(2r-2Q_{7}+Q_{9}-
4S_{2})(4S_{1}+Q_{1}-e^{\nu}\nu')+\\
\vspace{0.35 cm} 
e^{\lambda}[4S_{5}(Q_{10}-S_{5}-2S_{7})-Q^{2}_{8}+2r^{2}(\dot{Q_{8}}-
2\dot{Q_{6}}-4\dot{S{4}}+2S_{4}\dot{\nu})+\\
\vspace{0.35 cm}  
Q_{8}(4S_{4}+2Q_{6}-r^{2}\dot{\nu})+
2r^{2}Q_{6}\dot{\nu}]\}-
\frac{1}{4r^{2}}e^{-2\nu}Q_{0}(2Q_{6}-Q_{8}+4S_{4}),\\
\vspace{0.35 cm}  
F_{3}=-\frac{1}{4r^{2}}e^{-\frac{1}{2}(\lambda+\nu)}[\frac{4}{r^{2}}S_{5}
(Q_{11}-S_{6}+2S_{8})-8S_{4}'+8S_{4}\nu'+2(Q_{8}'-2Q_{6}'+\\
\vspace{0.35 cm}  
2\nu'Q_{6}+\nu'Q_{8})]+\frac{1}{4r^{2}}e^{-\frac{3}{2}\lambda-\nu}
(2r+Q_{9}-2Q_{7}-
4S_{4})(Q_{4}-2Q_{3}-4S_{3}+\\
\vspace{0.35 cm}  
e^{\lambda}\dot{\lambda})-\frac{1}{4r^{2}}e^{-\frac{1}{2}\lambda-
\frac{3}{2}\nu}(2Q_{6}-
Q_{8}+4S_{4})[Q_{1}+e^{\nu}(\frac{2}{r}+\frac{1}{r^{2}}Q_{9}-\nu')],\\
\vspace{0.35 cm}  
F_{4}=\frac{1}{4}e^{-(\lambda+2\nu)}[Q^{2}_{1}+Q_{0}(Q_{4}-4S_{3}+
e^{\lambda}\dot{\lambda})+Q_{1}(4S_{1}-2Q_{2}-e^{\nu}\lambda'-\\
\vspace{0.35 cm}  
2e^{\nu}\nu')]+
\frac{1}{4}e^{-(2\lambda+\nu)}\{Q_{5}(Q_{1}+4S_{1})+Q^{2}_{4}+
e^{\lambda}[2\dot{Q_{4}}-
8\dot{S_{3}}+4S_{3}(\dot{\lambda}+\dot{\nu})+\\
\vspace{0.35 cm}  
e^{\lambda}(\dot{\lambda}^{2}+2\ddot{\lambda}-\dot{\lambda}\dot{\nu})-
Q_{4}\dot{\nu}+2Q_{1}'-4Q_{2}'+8S_{1}'+(2Q_{2}-4S_{1})(\lambda'+\\
\vspace{0.35 cm}  
\nu')+
e^{\nu}(\lambda'\nu'-\nu{'}^{2}-2\nu{'}{'})]-2Q_{2}Q_{5}-4Q_{4}S_{3}-
e^{\nu}Q_{5}\nu'\},\\
\vspace{0.35 cm}
\end{array}
\end{equation}
\begin{displaymath}
\begin{array}{c}
\vspace{0.35 cm}   
F_{5}=\frac{1}{2}e^{-\frac{3}{2}(\lambda+\nu)}[Q_{3}(e^{\nu}\nu'-Q_{1}-
4S_{1})+e^{\nu}(Q_{5}\dot{\lambda}-\dot{Q_{5}}+Q_{4}'-Q_{4}\lambda')+\\
\vspace{0.35 cm} 
Q_{2}(2Q_{3}-Q_{4}+4S_{3}-e^{\lambda}\dot{\lambda})],\\
\vspace{0.35 cm}   
F_{6}=-\frac{1}{4r^{4}}e^{-\frac{1}{2}(\lambda+\nu)}[(-r^{2}e^{-\lambda}
Q_{4}+Q_{8}+r^{2}\dot{\lambda})(2r-2Q_{7}+\\
\vspace{0.35 cm}
Q_{9}-4S_{2})+S_{6}(-4Q_{10}+4S_{5}+
8S_{7})+r^{2}(4\dot{Q_{7}}-2\dot{Q_{9}}+
8\dot{S_{2}})+\\
\vspace{0.35 cm}
r^{2}(2Q_{6}-Q_{8}+4S_{4})(-\nu'+e^{-\nu}Q_{1}+
4e^{-\nu}S_{1}-2e^{-\nu}Q_{2})],\\
\vspace{0.35 cm}
F_{7}=-\frac{1}{4r^{4}}[-e^{-2\lambda}r^{2}Q_{5}(2r-2Q_{7}+Q_{9}-4S_{2})+
e^{-(\lambda+\nu)}r^{2}(2Q_{6}-Q_{8}+4S_{4})\\
\vspace{0.35 cm} 
(4S_{3}-Q{4}-e^{\lambda}\dot{\lambda})]-
\frac{1}{4r^{4}}e^{-\lambda}[Q_{9}(4r+
Q_{9}-4S_{2})-8rS_{2}+4S_{6}(S_{6}-Q_{11}+\\
\vspace{0.35 cm} 
2S_{8})+2r^{2}(2Q_{7}'-Q_{9}'+4S_{2}'+
\lambda'r+\frac{1}{2}\lambda'Q_{9}-2\lambda'S_{2})-
2Q_{7}(Q_{9}+2r+r^{2}\lambda')],\\
\vspace{0.35 cm} 
F_{8}=\frac{1}{4r^{4}}e^{-\nu}[(Q_{8}-4S_{4})(Q_{8}+e^{-\nu}r^{2}
Q_{0}-r^{2}\dot{\nu})-4(Q_{10}-S_{5})(Q_{10}-S_{5}-\\
\vspace{0.35 cm} 
2S_{7})+2r^{2}(\dot{Q_{8}}-4\dot{S_{4}})+e^{-\lambda}r^{2}(2r+Q_{9}-
4S_{2})(Q_{1}-2Q_{2}+4S_{1}-e^{\nu}\nu')],\\
\vspace{0.35 cm} 
F_{9}=\frac{1}{4r^{4}}e^{-\frac{1}{2}(\lambda+\nu)}[(Q_{8}-4S_{4})(Q_{9}-
2r+e^{-\nu}r^{2}Q_{1}-r^{2}\nu')-4(Q_{10}-S_{5})
(Q_{11}-\\
\vspace{0.35 cm} 
S_{6}-2S_{8})-e^{-\lambda}r^{2}(2r+Q_{9}-4S_{2})(Q_{4}-
4S_{3}+e^{\lambda}\dot{\lambda})+2r^{2}(Q_{8}'-4S_{4}')],\\
\vspace{0.35 cm} 
F_{10}=-\frac{1}{4r^{4}}e^{-\frac{1}{2}(\lambda+\nu)}\{(2r+Q_{9}-4S_{4})
[Q_{8}-e^{-\lambda}r^{2}(Q_{4}+e^{\lambda}\dot{\lambda})]-4(Q_{11}-\\
\vspace{0.35 cm} 
S_{6})(Q_{10}-S_{5}-2S_{7})+2r^{2}(\dot{Q_{9}}-4\dot{S_{2}})+
e^{-\nu}r^{2}(Q_{8}-4S_{4})(Q_{1}+4S_{1}-e^{\nu}\nu')\},\\
\vspace{0.35 cm} 
F_{11}=-\frac{1}{4r^{4}}e^{-\lambda}[4r^{2}-4(Q_{11}-S_{6})(Q_{11}-S_{6}-
2S_{8})+\\
\vspace{0.35 cm}
e^{-\nu}r^{2}(Q_{8}-
4S_{4})(2Q_{3}- 
Q_{4}+4S_{3}-e^{\lambda}\dot{\lambda})+2r^{2}(Q_{9}'-4S_{2}')+\\
\vspace{0.35 cm}
(2r+Q_{9}-4S_{2})
(Q_{9}-2r-r^{2}e^{-\lambda}Q_{5}-r^{2}\lambda')],\\
\vspace{0.35 cm} 
F_{12}=-\frac{1}{2r^{3}}e^{-\frac{1}{2}(\lambda+\nu)}[2Q_{8}+
r(\dot{Q_{9}}-Q_{8}')].\\
\vspace{0.35 cm} 
F_{13}=-\frac{1}{r^{2}}+\frac{1}{4r^{4}}e^{-\lambda}[(2r+Q_{9}-
4S_{2})(2r-2Q_{7}+Q_{9}-
4S_{2})+4S_{6}(S_{6}-\\
\vspace{0.35 cm} 
Q_{11})]-
\frac{1}{4r^{4}}e^{-\nu}[(Q_{8}-4S_{4})(Q_{8}-2Q_{6}-
4S_{4})+4S_{5}(S_{5}-Q_{10})],\\
\vspace{0.35 cm} 
F_{14}=\frac{1}{r^{4}}e^{-\nu}[Q_{10}(Q_{8}-2Q_{6}-4S_{4})+2Q_{6}S_{5}],\\
\vspace{0.35 cm}  
F_{15}=-\frac{1}{r^{4}}e^{-\frac{1}{2}(\lambda+\nu)}[(Q_{11}-S_{6})(Q_{8}-
2Q_{6}-4S_{4})+S_{5}(2r+Q_{9}-4S_{2})],\\
\vspace{0.35 cm} 
F_{16}=-\frac{1}{r^{4}}e^{-\frac{1}{2}(\lambda+\nu)}[(Q_{10}-S_{5})(2r-2Q_{7}
+Q_{9}-4S_{2})+S_{6}(Q_{8}-4S_{4})],\\
\vspace{0.35 cm} 
F_{17}=\frac{1}{r^{4}}e^{-\lambda}[Q_{11}(2r-2Q_{7}+Q_{9}-4S_{2})+
2Q_{7}S_{6}],\\
\vspace{0.35 cm}
\end{array}
\end{displaymath}
\begin{displaymath}
\begin{array}{c}
\vspace{0.35 cm} 
F_{18}=\frac{1}{2r^{4}}e^{-\nu}\{(Q_{8}-4S_{4})(Q_{10}-S_{5}-2S_{7})+
2r^{2}(\dot{Q_{10}}-\dot{S_{5}})+(Q_{10}-\\
\vspace{0.35 cm} 
S_{5})[Q_{8}+
e^{-\nu}r^{2}(Q_{0}-e^{\nu}\dot{\nu})]+e^{-\lambda}r^{2}(Q_{11}-
S_{6})(Q_{1}-2Q_{2}+4S_{1}-e^{\nu}\nu')\},\\
\vspace{0.35 cm} 
F_{19}=\frac{1}{2r^{4}}e^{-\frac{1}{2}(\lambda+\nu)}\{(Q_{10}-S_{5})
[Q_{9}-2r+e^{-\nu}r^{2}(Q_{1}-e^{\nu}\nu')]+(Q_{8}-4S_{4})\\
\vspace{0.35 cm} 
(Q_{11}-
S_{6}-2S_{8})-e^{-\lambda}r^{2}(Q_{11}-S_{6})(Q_{4}-
4S_{3}+e^{\lambda}\dot{\lambda})+2r^{2}(Q_{10}'-S_{5}')\},\\
\vspace{0.35 cm} 
F_{20}=-\frac{1}{2r^{4}}\{e^{-2\nu}r^{2}Q_{0}S_{5}-e^{-(\lambda+\nu)}
r^{2}S_{6}(Q_{1}+4S_{1})+
e^{-\nu}[-Q_{10}(2Q_{6}-Q_{8}+\\
\vspace{0.35 cm} 
4S_{4})-
2Q_{8}S_{7}+(S_{5}+2S_{7})(4S_{4}+
2Q_{6})-2r^{2}\dot{S_{5}}+r^{2}S_{5}\dot{\nu}]+
e^{-\lambda}r^{2}S_{6}\nu'\},\\
\vspace{0.35 cm} 
F_{21}=\frac{1}{2r^{4}}e^{-\frac{3}{2}\lambda-
\frac{1}{2}\nu}S_{6}[r^{2}(2Q_{3}-Q_{4}+4S_{3})+e^{\lambda}(Q_{8}-2Q_{6}-
4S_{4}-r^{2}\dot{\lambda})]+\\
\vspace{0.35 cm}  
\frac{1}{r^{4}}e^{-\frac{1}{2}(\lambda+\nu)}[\frac{1}{2}Q_{11}(2Q_{6}-
Q_{8}+4S_{4})+S_{8}(Q_{8}-
2Q_{6}-4S_{4})+r^{2}S_{5}']-\\
\vspace{0.35 cm} 
\frac{1}{2r^{4}}e^{-\frac{1}{2}\lambda-\frac{3}{2}\nu}S_{5}[r^{2}Q_{1}+
e^{\nu}(Q_{9}+2r+r^{2}\nu')],\\
\vspace{0.35 cm} 
F_{22}=-\frac{1}{2r^{4}}e^{-\frac{1}{2}(\lambda+\nu)}[(Q_{11}-S_{6})
(e^{-\lambda}r^{2}Q_{4}+r^{2}\dot{\lambda}-
Q_{8})-(2r+Q_{9}-4S_{2})(Q_{10}-\\
\vspace{0.35 cm} 
S_{5}-2S_{7})-2r^{2}(\dot{Q_{11}}-\dot{S_{6}})+
e^{-\nu}r^{2}(Q_{10}-S_{5})(-Q_{1}-4S_{1}+e^{\nu}\nu')],\\
\vspace{0.35 cm}   
F_{23}=-\frac{1}{2r^{4}}e^{-\lambda}[(Q_{11}-S_{6})(2r-Q_{9}+
e^{-\lambda}r^{2}Q_{5}+r^{2}\lambda')-(2r+Q_{9}-4S_{2})(Q_{11}-\\
\vspace{0.35 cm} 
S_{6}-2S_{8})-e^{-\nu}r^{2}(Q_{10}-S_{5})(2Q_{9}-
Q_{4}+4S_{3}-e^{\lambda}\dot{\lambda})-2r^{2}(Q_{11}'-S_{6}')],\\
\vspace{0.35 cm}   
F_{24}=\frac{1}{2r^{4}}e^{-2\lambda-\nu}\{e^{\lambda}r^{2}S_{5}(Q_{4}-
4S_{3})-
r^{2}e^{\nu}Q_{5}S_{6}+e^{\lambda+\nu}[2S_{6}(Q_{7}+\\
\vspace{0.35 cm} 
2S_{2})+(Q_{11}-2S_{8})(2r-2Q_{7}+Q_{9}-4S_{2})+
r^{2}S_{6}\lambda'-2r^{2}S_{6}']+e^{2\lambda}r^{2}S_{5}\dot{\lambda}\},\\
\vspace{0.35 cm} 
F_{25}=\frac{1}{2r^{4}}e^{-\frac{1}{2}(\lambda+\nu)}[Q_{8}S_{6}+
(2r-2Q_{7}+Q_{9}-4S_{2})(Q_{10}-2S_{7}-e^{\nu}S_{5})+\\
\vspace{0.35 cm} 
r^{2}(S_{6}\dot{\lambda}-2\dot{S_{6}})+r^{2}S_{5}(2Q_{2}-Q_{1}-4S_{1}+
e^{\nu}\nu')]-\frac{1}{2r^{2}}e^{-\frac{1}{2}(3\lambda+\nu)}Q_{4}S_{6},\\
\vspace{0.35 cm} 
F_{26}=\frac{1}{r^{3}}e^{-\frac{1}{2}(\lambda+\nu)}[2Q_{10}-2S_{5}-4S_{7}+
r(\dot{S_{6}}-2\dot{S_{8}}-Q_{10}'+S_{5}'+2S_{7}')].\\
\vspace{0.35 cm} 
\end{array}
\end{displaymath}

Let us consider Bianchi identities, which can be written in the following 
form: 
 
\begin{equation}
\varepsilon^{\sigma\lambda\mu\nu}\nabla_{\lambda}F^{i}{}_{k\mu\nu}=0.
\end{equation}
where $\nabla$ is differential operator defined analogously to covariant 
derivative determinet by means of Cristoffel symbols or connection 
$(-A^{ik}{}_{\mu})$ in case of holonomic and anholonomic indices 
respectively:
 
$\nabla_{\lambda}F^{i}{}_{k\mu\nu}=
\partial_{\lambda}F^{i}{}_{k\mu\nu}+A^{l}{}_{k\lambda}F^{i}{}_{l\mu\nu}-
A^{i}{}_{l\lambda}F^{l}{}_{k\mu\nu}-{\left\{{\sigma 
\hfill}\atop{\mu\lambda}\right\}}F^{i}{}_{k\sigma\nu}-
{\left\{{\sigma\hfill}\atop{\nu\lambda}\right\}}F^{i}{}_{k\mu\sigma}$, 
and $\varepsilon^{\sigma\lambda\mu\nu}$ is Levi-Chivita symbol. 

 In spherically-symmetric case Bianchi identities are reduced to 20 
relations: 

\begin{displaymath}
\begin{array}{c} 
\vspace{0.35 cm} 
e^{\frac{1}{2}(\lambda+\nu)}(4F_{14}+2r^{2}F_{14}')+e^{\lambda}[2F_{19}
(2Q_{6}-Q_{8}+4S_{4})-\\
\vspace{0.35 cm} 
2F_{21}(Q_{8}-4S_{4})+4F_{3}(Q_{10}-S_{5})-4F_{9}S_{5}+
r^{2}\dot{\lambda}(F_{15}+F_{16})]+\\
\vspace{0.35 cm} 
r^{2}[(F_{15}+F_{16})(Q_{4}-4S_{3})-2F_{16}Q_{3}]=0,\\
\vspace{0.35 cm} 
-rF_{15}(e^{\lambda}rQ_{1}+e^{\nu}rQ_{5})+e^{\lambda+\nu}[2F_{21}(2r+Q_{9}-
4S_{2})+4rF_{15}-4F_{3}(Q_{11}-\\
\vspace{0.35 cm} 
S_{6})+2r^{2}F_{15}']+e^{\frac{1}{2}(\lambda+\nu)}
[-e^{\lambda}(2F_{23}(2Q_{6}-Q_{8}+4S_{4})+\\
\vspace{0.35 cm} 
4F_{11}S_{5})+r^{2}(F_{17}+F_{14})(Q_{4}-2Q_{3}-4S_{4}+
e^{\lambda}\dot{\lambda})]=0,\\
\vspace{0.35 cm} 
rF_{16}(e^{\lambda}rQ_{1}+e^{\nu}rQ_{5})+2e^{\lambda+\nu}[2rF_{16}+F_{19}(2r-
2Q_{7}+Q_{9}-4S_{2})+\\
\vspace{0.35 cm} 
2F_{9}S_{6}+r^{2}F_{16}']+e^{\frac{1}{2}(\lambda+\nu)}[r^{2}(F_{17}+
F_{14})(Q_{4}-4S_{3}+e^{\lambda}\dot{\lambda})+\\
\vspace{0.35 cm} 
e^{\lambda}F_{24}(8S_{4}-2Q_{8}+4Q_{10}-4S_{5})]=0,\\
\vspace{0.35 cm} 
4rF_{17}+2(F_{24}-F_{23})(2r-2Q_{7}+Q_{9}-4S_{2})+4F_{24}Q_{7}-4F_{7}(Q_{11}-
S_{6})+\\
\vspace{0.35 cm} 
4F_{11}S_{6}+e^{-\frac{1}{2}(\lambda+\nu)}r^{2}[(F_{15}+F_{16})(Q_{4}-
4S_{3}+e^{\lambda}\dot{\lambda})-2F_{16}Q_{3})]+2r^{2}F_{17}'=0,\\
\vspace{0.35 cm}  
4r(F_{14}-F_{17})-4F_{23}Q_{7}+2(F_{23}-F_{24})(2r+Q_{9}-4S_{2})+\\
\vspace{0.35 cm} 
e^{\frac{1}{2}(\lambda-\nu)}[4F_{19}Q_{6}+2(F_{19}+F_{21})(-Q_{8}+4S_{4})+\\
\vspace{0.35 cm} 
4F_{3}(Q_{10}-S_{5})-
4F_{9}S_{5}]+4F_{7}(Q_{11}-S_{6})-4F_{11}+2r^{2}(F_{14}'-F_{17}')=0,\\
\vspace{0.35 cm}  
4rF_{13}+(F_{11}-F_{7})(2r+Q_{9}-4S_{2})-2F_{11}Q_{7}+\\
\vspace{0.35 cm} 
e^{\frac{1}{2}(\lambda-\nu)}[(Q_{8}-4S_{4})(F_{9}-F_{3})-2F_{9}Q_{6}-
2F_{21}(Q_{10}-S_{5})-2F_{19}S_{5}]-\\
\vspace{0.35 cm} 
2F_{24}(Q_{11}-S_{6})+2F_{23}S_{6}+2r^{2}F_{13}'=0,\\
\vspace{0.35 cm}  
2e^{\frac{1}{2}(\lambda+\nu)}[(4S_{4}-Q_{8})(F_{18}+F_{20})+2Q_{6}F_{18}+
2F_{2}(Q_{10}-S_{5})-\\
\vspace{0.35 cm}
2F_{8}S_{5}+ 
r^{2}\dot{F_{14}}]-r^{2}(F_{15}+F_{16})(Q_{1}+
4S_{1}-e^{\nu}\nu')+2r^{2}F_{15}Q_{2}=0,\\
\vspace{0.35 cm} 
-r^{2}F_{15}(e^{\lambda}Q_{0}+e^{\nu}Q_{4})+2e^{\frac{1}{2}(\lambda+3\nu)}
[F_{20}(2r+Q_{9}-4S_{2})-2F_{2}(Q_{11}-S_{6})]-\\
\vspace{0.35 cm} 
2e^{\lambda+\nu}[F_{22}(2Q_{6}-Q_{8}+4S_{4})+2F_{10}S_{5}-
r^{2}\dot{F_{15}}]-\\
\vspace{0.35 cm} 
e^{\frac{1}{2}(\lambda+\nu)}r^{2}(F_{14}+F_{17})(Q_{1}+4S_{1}-
e^{\nu}\nu')=0,\\
\vspace{0.35 cm}
\end{array}
\end{displaymath}
\begin{equation}
\begin{array}{c}
\vspace{0.35 cm}      
r^{2}F_{16}(e^{\lambda}Q_{0}+e^{\nu}Q_{4})+2e^{\frac{1}{2}(\lambda+3\nu)}
[F_{18}(2r-2Q_{7}+Q_{9}-4S_{2})+F_{8}S_{6}]+\\
\vspace{0.35 cm} 
2e^{\lambda+\nu}[-F_{25}(Q_{8}-4S_{4})+2F_{6}(Q_{10}-S_{5})+
r^{2}\dot{F_{16}}]-\\
\vspace{0.35 cm} 
e^{\frac{1}{2}(\lambda+\nu)}r^{2}(F_{14}+
F_{17})(Q_{1}-2Q_{2}+4S_{1}-e^{\nu}\nu')=0,\\
\vspace{0.35 cm} 
e^{\nu}[4F_{6}(S_{6}-Q_{11})+2(F_{25}-F_{22})(2r+Q_{9}-4S_{2})+
4F_{22}Q_{7}+\\
\vspace{0.35 cm}
4F_{10}S_{6}+ 
r^{2}\nu'(F_{15}+F_{16})]+
r^{2}[2F_{15}Q_{2}-\\
\vspace{0.35 cm}
(Q_{1}+4S_{1})(F_{15}+
F_{16})]+2e^{\frac{1}{2}(\lambda+\nu)}r^{2}\dot{F_{17}}=0,\\
\vspace{0.35 cm}  
e^{\frac{1}{2}\lambda}[2F_{2}(Q_{10}-S_{5})+F_{18}(2Q_{6}-Q_{8}+4S_{4})+
F_{20}(4S_{4}-Q_{8})-\\
\vspace{0.35 cm}
2F_{8}S_{5}+r^{2}(\dot{F_{14}}- 
\dot{F_{17}})]+e^{\frac{1}{2}\nu}
[2F_{6}(Q_{11}-S_{6})+\\
\vspace{0.35 cm}
(F_{22}-
F_{25})(2r+Q_{9}-4S_{2})-2Q_{7}F_{22}-2F_{10}S_{6}]=0,\\
\vspace{0.35 cm}  
e^{\frac{1}{2}(\nu-\lambda)}[(F_{6}-F_{10})(2r+Q_{9}-4S_{2})+2F_{10}Q_{7}+
2F_{25}(Q_{11}-S_{6})-2F_{22}S_{6}]+\\
\vspace{0.35 cm} 
(F_{2}-F_{8})(Q_{8}-4S_{4})+2F_{8}Q_{6}+2F_{20}(Q_{10}-S_{5})+2F_{18}S_{5}-
2r^{2}\dot{F_{13}}=0,\\
\vspace{0.35 cm}  
e^{\frac{1}{2}(\lambda+3\nu)}[F_{2}(-Q_{9}+2r+r^{2}\nu')+
F_{1}(2r-2Q_{7}+Q_{9}-4S_{2})-4F_{20}(Q_{11}-\\
\vspace{0.35 cm}
S_{6}-2S_{8})+4r^{2}F_{2}']+
2e^{\frac{1}{2}(\lambda+\nu)}r^{2}[F_{7}(Q_{1}+4S_{1}-e^{\nu}\nu')-
F_{2}Q_{1}]+\\
\vspace{0.35 cm} 
2e^{\lambda}r^{2}F_{3}Q_{0}+2e^{\lambda+\nu}[F_{3}(Q_{8}-
r^{2}\dot{\lambda})+(F_{0}+F_{12})(2Q_{6}-Q_{8}+4S_{4})-
2F_{26}S_{5}+\\
\vspace{0.35 cm} 
2F_{21}(Q_{10}-S_{5}-2S_{7})+r^{2}(-2\dot{F_{3}}+
F_{6}\dot{\lambda})]-2e^{\nu}r^{2}F_{6}(2Q_{3}-Q_{4}+4S_{3})=0,\\
\vspace{0.35 cm}   
\frac{1}{r}e^{\frac{1}{2}\nu}[F_{20}(-2r+Q_{9})+F_{1}S_{6}-2F_{2}(Q_{11}-
S_{6}+2S_{8})-r^{2}(2F_{20}'+\\
\vspace{0.35 cm} 
F_{20}\nu')]-e^{\frac{1}{2}\lambda-\nu}rF_{21}Q_{0}+
e^{-\frac{1}{2}\nu}r[F_{20}Q_{1}+F_{24}(-Q_{1}+4S_{1}+e^{\nu}\nu')]+\\
\vspace{0.35 cm} 
\frac{1}{r}e^{\frac{1}{2}\lambda}[-F_{21}Q_{8}+F_{26}(-2Q_{6}+Q_{8}-4S_{4})-
2S_{5}(F_{0}+F_{12})+2F_{3}(Q_{10}-\\
\vspace{0.35 cm} 
S_{5}-2S_{7})+r^{2}(2\dot{F_{21}}+F_{21}\dot{\lambda})]-
e^{-\frac{1}{2}\lambda}
rF_{25}(-2Q_{3}+Q_{4}-4S_{3}+e^{\lambda}\dot{\lambda})=0,\\
\vspace{0.35 cm}    
e^{\lambda+\nu}[(F_{12}-F_{5})(2r-2Q_{7}+Q_{9}-4S_{2})+2F_{26}S_{6}-
2F_{25}(Q_{11}-S_{6}-2S_{8})+\\
\vspace{0.35 cm} 
2r^{2}F_{6}'+F_{6}(-Q_{9}+2r+r^{2}\nu')]+e^{\frac{1}{2}(\lambda+\nu)}
r^{2}[Q_{4}(F_{2}-F_{7})-4F_{2}S_{3}]+\\
\vspace{0.35 cm} 
e^{\frac{1}{2}(3\lambda+\nu)}[2F_{4}(Q_{6}+2S_{4})+Q_{8}(F_{7}-F_{4})+
2F_{24}(Q_{10}-S_{5}-2S_{7})+
r^{2}(-2\dot{F_{7}}+\\
\vspace{0.35 cm} 
\dot{\lambda}F_{2}-\dot{\lambda}F_{7})]+e^{\lambda}r^{2}F_{3}
(Q_{1}-
2Q_{2}+4S_{1}-e^{\nu}\nu')+e^{\nu}r^{2}F_{6}Q_{5}=0,\\
\vspace{0.35 cm}
\end{array}
\end{equation}
\begin{displaymath}
\begin{array}{c}
\vspace{0.35 cm}  
-e^{\lambda+\nu}[F_{26}(2r-2Q_{7}+Q_{9}-4S_{2})-2S_{6}(F_{12}-F_{5})+
2F_{6}(Q_{11}-S_{6}-\\
\vspace{0.35 cm} 
2S_{8})+2r^{2}F_{25}'+F_{25}(-Q_{9}+2r+r^{2}\nu')]-e^{\frac{1}{2}(\lambda+\nu)}
r^{2}[Q_{4}(F_{20}-F_{24})-\\
\vspace{0.35 cm} 
4F_{20}S_{3}]-e^{\frac{1}{2}(3\lambda+\nu)}[F_{24}Q_{8}+
2F_{4}S_{5}-2F_{7}(Q_{10}-S_{5}-2S_{7})+r^{2}(-2\dot{F_{24}}+\\
\vspace{0.35 cm} 
\dot{\lambda}F_{20}-\dot{\lambda}F_{24})]-e^{\lambda}r^{2}F_{21}(Q_{1}-2Q_{2}+4S_{1}-
e^{\nu}\nu')-e^{\nu}r^{2}F_{25}Q_{5}=0,\\
\vspace{0.35 cm}     
e^{\frac{1}{2}\nu}[-2F_{8}-2rF_{8}'-r\nu'(F_{8}-F_{11})]+e^{-\frac{1}{2}\nu}
r[-F_{8}Q_{1}+F_{11}(-Q_{1}+2Q_{2}-\\
\vspace{0.35 cm} 
4S_{1})]+e^{\frac{1}{2}\lambda-\nu}rF_{9}Q_{0}+
e^{\frac{1}{2}\lambda}r[2\dot{F_{9}}+
\dot{\lambda}(F_{9}-F_{10})]-e^{-\frac{1}{2}\lambda}rF_{10}(Q_{4}-4S_{3})+\\
\vspace{0.35 cm} 
\frac{1}{r}e^{\frac{1}{2}\lambda}[F_{9}Q_{8}+(F_{0}+F_{12})(Q_{8}-4S_{4})+
2(F_{26}-F_{19})(Q_{10}-S_{5})+4F_{19}S_{7}]+\\
\vspace{0.35 cm} 
\frac{1}{r}e^{\frac{1}{2}\nu}[-F_{8}Q_{9}+F_{4}(2r+Q_{9}-4S_{2})+2F_{18}
(Q_{11}-S_{6}-2S_{8})]=0,\\
\vspace{0.35 cm}   
e^{\frac{1}{2}\nu}[2F_{10}+2rF_{10}'+r\nu'(F_{10}-F_{9})]+e^{-\frac{1}{2}\nu}
rF_{9}(Q_{1}+4S_{4})+\\
\vspace{0.35 cm}
e^{-\frac{1}{2}\lambda}r[F_{11}Q_{4}+F_{8}(-2Q_{3}+Q_{4}-4S_{3})]-
e^{\frac{1}{2}\lambda}r[2\dot{F_{11}}+\dot{\lambda}(F_{11}-F_{8})]-\\
\vspace{0.35 cm}
e^{-\lambda+\frac{1}{2}\nu}rF_{10}Q_{5}-
\frac{1}{r}e^{\frac{1}{2}\lambda}
[F_{11}Q_{8}+\frac{1}{2}F_{1}(Q_{8}-
4S_{4})-2F_{23}(Q_{10}-\\
\vspace{0.35 cm}
S_{5}-2S_{7})]+
\frac{1}{r}e^{\frac{1}{2}\nu}[F_{10}Q_{9}+(F_{12}-
F_{5})(2r+Q_{9}-4S_{2})+\\
\vspace{0.35 cm}
2(F_{26}+F_{22})(Q_{11}-S_{6})-4F_{22}S_{8}]=0,\\
\vspace{0.35 cm}
-e^{\frac{1}{2}\lambda-\nu}rF_{19}Q_{0}+
e^{\frac{1}{2}\nu}[2F_{18}+2rF_{18}'+r\nu'(F_{18}+F_{23})]+
e^{-\frac{1}{2}\nu}r[F_{18}Q_{1}+\\
\vspace{0.35 cm}
F_{23}(-Q_{1}+Q_{2}-4S_{1})]-
e^{-\frac{1}{2}\lambda}rF_{22}(Q_{4}-4S_{3})-
e^{\frac{1}{2}\lambda}r[2\dot{F_{19}}+\dot{\lambda}(F_{19}+F_{22})]-\\
\vspace{0.35 cm}
\frac{1}{r}e^{\frac{1}{2}\lambda}[F_{19}Q_{8}+F_{26}(Q_{8}-4S_{4})-
2(F_{0}+F_{12}+F_{9})(Q_{10}-S_{5})+4F_{9}S_{7}]+\\
\vspace{0.35 cm}
\frac{1}{r}e^{\frac{1}{2}\nu}[F_{18}Q_{9}+
2(F_{4}-F_{8})(-Q_{11}+S_{6})-4F_{8}S_{8}]=0,\\
\vspace{0.35 cm}
-e^{\frac{1}{2}\nu-\lambda}rF_{22}Q_{5}+
e^{\frac{1}{2}\nu}[2F_{22}+2rF_{22}'+r\nu'(F_{22}+F_{19})]-
e^{-\frac{1}{2}\nu}rF_{19}(Q_{1}+4S_{1})+\\
\vspace{0.35 cm}
e^{-\frac{1}{2}\lambda}r[F_{23}Q_{4}-F_{18}(-2Q_{3}+Q_{4}-4S_{3})]-
e^{\frac{1}{2}\lambda}r[2\dot{F_{23}}+\dot{\lambda}(F_{23}+F_{18})]-\\
\vspace{0.35 cm}
\frac{1}{r}e^{\frac{1}{2}\lambda}[F_{23}Q_{8}-(F_{1}-2F_{11})(Q_{10}-
S_{5})-4F_{11}S_{7}]+\frac{1}{r}e^{\frac{1}{2}\nu}[F_{22}Q_{9}+\\
\vspace{0.35 cm}
F_{26}(2r+Q_{9}-4S_{2})+
2(F_{5}-F_{12}-F_{10})(Q_{11}-S_{6})+4F_{10}S_{8}]=0.\\
\vspace{0.35 cm}
\end{array}
\end{displaymath}
 In the case of vanishing pseudoscalar functions $S_{i}$ ($i=5,\ \ldots 8$)
and $Q_{i}(i = 10, 11)$ the curvature tensor is determined by 14 functions
$\tilde{F}_{i}(i = 0, 1, \ldots 13)$
(othes functions $F_{i}(i = 14, \ldots 26)$ are equal to zero), 
expressions of which follow from (9): 
\begin{equation}
\begin{array}{c}
\vspace{0.35 cm} 
\ft{0}{0}{1}{0}=\tilde{F}_{0},\ 
\ft{0}{1}{1}{0}=\tilde{F}_{1},\ 
\ft{0}{2}{2}{0}=\ft{0}{3}{3}{0}=\tilde{F}_{2},\ 
\ft{0}{2}{2}{1}=\ft{0}{3}{3}{1}=\tilde{F}_{3},\\
\vspace{0.35 cm}
\ft{1}{0}{1}{0}=\tilde{F}_{4},\ \ft{1}{1}{1}{0}=\tilde{F}_{5},\ 
\ft{1}{2}{2}{0}=\ft{1}{3}{3}{0}=\tilde{F}_{6},\ 
\ft{1}{2}{2}{1}=\ft{1}{3}{3}{1}=\tilde{F}_{7},\\
\vspace{0.35 cm}
\ft{2}{0}{2}{0}=\ft{3}{0}{3}{0}=\tilde{F}_{8},\
\ft{2}{0}{2}{1}=\ft{3}{0}{3}{1}=\tilde{F}_{9},\
\ft{2}{1}{2}{0}=\ft{3}{1}{3}{0}=\tilde{F}_{10},\\
\vspace{0.35 cm}
\ft{2}{1}{2}{1}=\ft{3}{1}{3}{1}=\tilde{F}_{11},\
\ft{2}{2}{0}{1}=\ft{3}{3}{0}{1}=\tilde{F}_{12},\
-\ft{3}{2}{3}{2}=\ft{2}{3}{3}{2}=\tilde{F}_{13},\\
\vspace{0.35 cm}
\end{array}
\end{equation}
where explicit form of functions $\tilde{F}_{i}$ is:
\begin{equation}
\begin{array}{c}
\vspace{0.35 cm}
\tilde{F}_{0}=\frac{1}{2}\{e^{-\frac{3}{2}(\lambda+\nu)}[-Q_{3}(4S_{1}+
Q_{1})+
Q_{2}(2Q_{3}-Q_{4}+4S_{3})]+\\
\vspace{0.35 cm}
e^{-\frac{1}{2}(\lambda+3\nu)}(-\dot{Q_{1}}-
Q_{2}\dot{\lambda}+Q_{1}\dot{\nu}+
Q_{0}'-Q_{0}\nu')+e^{-\frac{1}{2}(3\lambda+\nu)}Q_{3}\nu'\},\\
\vspace{0.35 cm}
\tilde{F}_{1}=\frac{1}{4}\{e^{-\lambda-2\nu}[Q_{1}^{2}+Q_{0}(-2Q_{3}+Q_{4}-
4S_{3}+e^{\lambda}\dot{\lambda})+4Q_{1}S_{1}]+\\
\vspace{0.35 cm}
e^{-2\lambda-\nu}[Q_{5}(Q_{1}+4S_{1}-e^{\nu}\nu')+Q_{4}(Q_{4}-
2Q_{3}-4S_{3})]+e^{-(\lambda+\nu)}[4\dot{Q_{3}}-\\
\vspace{0.35 cm}
2\dot{Q_{4}}+8\dot{S_{3}}+Q_{1}\lambda'+
(\dot{\lambda}+\dot{\nu})(Q_{4}-4S_{3}-2Q_{3})+Q_{4}\dot{\lambda}-2Q_{1}'-\\
\vspace{0.35 cm}
8S_{1}'+4S_{1}(\lambda'+\nu')+e^{\lambda}
(-\dot{\lambda^{2}}+\dot{\lambda}\dot{\nu}-2\ddot{\lambda})]-
e^{-\lambda}(-\lambda'\nu'+\nu'^{2}+2\nu'')\},\\
\vspace{0.35 cm}
\tilde{F}_{2}=-\frac{1}{4r^{4}}\{e^{-(\lambda+\nu)}r^{2}(2r-2Q_{7}+Q_{9}-
4S_{2})(Q_{1}+4S_{1}-e^{\nu}\nu')+\\
\vspace{0.35 cm}
e^{-\nu}[(-Q_{8}+4S_{4}+2Q_{6})(Q_{8}+r^{2}\dot{\nu})
+r^{2}(-4\dot{Q_{6}}+2\dot{Q_{8}}-\\
\vspace{0.35 cm}
8\dot{S_{4}})]+e^{-2\nu}r^{2}Q_{0}(2Q_{6}-Q_{8}+4S_{4})\},\\
\vspace{0.35 cm}
\tilde{F}_{3}=\frac{1}{4r^{2}}\{e^{-\frac{1}{2}(\lambda+\nu)}
[-\frac{1}{r^{2}}(2r+Q_{9})(2Q_{6}-Q_{8}+4S_{4})+8S_{4}'-8S_{4}\nu'+
4Q_{6}'-\\
\vspace{0.35 cm}
2Q_{8}'-4\nu'Q_{6}+2\nu'Q_{8}]+
e^{-\frac{1}{2}(3\lambda+\nu)}(2r-2Q_{7}+Q_{9}-4S_{2})(-2Q_{3}+\\
\vspace{0.35 cm}
Q_{4}-4S_{1}+e^{\lambda}\dot{\lambda})+
e^{-\frac{1}{2}(\lambda+3\nu)}(2Q_{6}-Q_{8}+4S_{4})(-Q_{1}+e^{\nu}\nu')\},\\
\vspace{0.35 cm}
\tilde{F}_{4}=\frac{1}{4}\{e^{-\lambda-2\nu}[Q_{0}(Q_{4}-4S_{3}+e^{\lambda}
\dot{\lambda})+Q_{1}(Q_{1}-2Q_{2}+4S_{1}-e^{\nu}\lambda'-\\
\vspace{0.35 cm}
2e^{\nu}\nu')]+e^{-2\lambda-\nu}[Q_{4}(Q_{4}-
4S_{3})+Q_{5}(Q_{1}+4S_{1}-2Q_{2})]+e^{-\lambda-\nu}[2\dot{Q_{4}}-\\
\vspace{0.35 cm}
8\dot{S_{3}}+4S_{3}(\dot{\lambda}+\dot{\nu})-
Q_{4}\dot{\nu}+2Q_{1}'-4Q_{2}'+8S_{1}'+(\lambda'+\nu')(2Q_{2}-4S_{1})]+\\
\vspace{0.35 cm}
e^{-\nu}(\dot{\lambda}^{2}-\dot{\lambda}\dot{\nu}+2\ddot{\lambda})-
e^{-2\lambda}Q_{5}\nu'+e^{-\lambda}(\lambda'\nu'-\nu'^{2}-2\nu'')\},\\
\vspace{0.35 cm}
\end{array}
\end{equation}
\begin{displaymath}
\begin{array}{c}
\vspace{0.35 cm}
\tilde{F}_{5}=\frac{1}{2}e^{-\frac{3}{2}(\lambda+\nu)}[-Q_{3}(Q_{1}+4S_{1})+
e^{\nu}(-\dot{Q_{5}}+Q_{5}\dot{\lambda}+Q_{4}'-\\
\vspace{0.35 cm}
Q_{4}\lambda'+Q_{3}\nu')+
Q_{2}(2Q_{3}-Q_{4}+4S_{3}-e^{\lambda}\dot{\lambda})],\\
\vspace{0.35 cm}
\tilde{F}_{6}=-\frac{1}{4r^{4}}\{r^{2}e^{-\frac{1}{2}(\lambda+3\nu)}(2Q_{6}-
Q_{8}+4S_{4})(Q_{1}+4S_{1}-\\
\vspace{0.35 cm}
e^{\nu}\nu'-2Q_{2})+
e^{-\frac{1}{2}(\lambda+\nu)}[(2r-
2Q_{7}+Q_{9}-4S_{2})(Q_{8}+r^{2}\dot{\lambda}-\\
\vspace{0.35 cm}
e^{-\lambda}r^{2}Q_{4})+r^{2}(4\dot{Q_{7}}-2\dot{Q_{9}}+
8\dot{S_{2}})]\},\\
\vspace{0.35 cm}
\tilde{F}_{7}=-\frac{1}{4r^{4}}\{-e^{-2\lambda}r^{2}(Q_{5}-e^{\lambda}
\lambda')(2r-2Q_{7}+Q_{9}-4S_{2})+\\
\vspace{0.35 cm}
e^{-\lambda-\nu}r^{2}(2Q_{6}-Q_{8}+4S_{4})(4S_{3}-Q_{4}-e^{\lambda}
\dot{\lambda})+e^{-\lambda}[Q_{9}(4r+Q_{9}-\\
\vspace{0.35 cm}
4S_{2})-8rS_{2}+r^{2}(4Q_{7}'-2Q_{9}'+8S_{2}')-2Q_{7}(2r+Q_{9})]\},\\
\vspace{0.35 cm}
\tilde{F}_{8}=\frac{1}{4r^{4}}\{e^{-\nu}[(Q_{8}+e^{-\nu}r^{2}Q_{0}-
r^{2}\dot{\nu})(Q_{8}-4S_{4})+2r^{2}(\dot{Q_{8}}-4\dot{S_{4}})]+\\
\vspace{0.35 cm}
e^{-\lambda-\nu}r^{2}(2r+Q_{9}-4S_{2})(Q_{1}-2Q_{2}+4S_{1}-e^{\nu}\nu')\},\\
\vspace{0.35 cm}
\tilde{F}_{9}=\frac{1}{4r^{4}}e^{-\frac{1}{2}(\lambda+\nu)}[(Q_{8}-4S_{4})
(Q_{9}-2r+e^{-\nu}r^{2}Q_{1}-r^{2}\nu')-\\
\vspace{0.35 cm}
e^{-\lambda}r^{2}(2r+Q_{9}-4S_{2})(Q_{4}-4S_{3}+e^{\lambda}\dot{\lambda})+
2r^{2}(Q_{8}'-4S_{4}')],\\
\vspace{0.35 cm}
\tilde{F}_{10}=-\frac{1}{4r^{4}}e^{-\frac{1}{2}(\lambda+\nu)}[(2r+Q_{9}-
4S_{2})(Q_{8}-e^{-\lambda}r^{2}Q_{4}-r^{2}\dot{\lambda})+\\
\vspace{0.35 cm}
2r^{2}(\dot{Q_{9}}-4\dot{S_{2}})+e^{-\nu}r^{2}(Q_{8}-4S_{4})(Q_{1}+4S_{1}-
e^{\nu}\nu')],\\
\vspace{0.35 cm}
\tilde{F}_{11}=-\frac{1}{4r^{4}}e^{-\lambda}\{r^{2}[4+e^{-\nu}(Q_{8}-4S_{4})
(2Q_{3}-Q_{4}+4S_{3}-e^{\lambda}\dot{\lambda})+\\
\vspace{0.35 cm}
2Q_{9}'-8S_{2}']+
(2r+Q_{9}-4S_{2})(Q_{9}-2r-e^{-\lambda}r^{2}Q_{5}+r^{2}\lambda')\},\\
\vspace{0.35 cm}
\tilde{F}_{12}=\frac{1}{2r^{3}}e^{-\frac{1}{2}(\lambda+\nu)}[2Q_{8}+
r(\dot{Q_{9}}-Q_{8}')],\\
\vspace{0.35 cm}
\tilde{F}_{13}=-\frac{1}{4r^{4}}[4r^{2}-e^{-\lambda}(2r+Q_{9}-4S_{2})
(2r-2Q_{7}+Q_{9}-4S_{2})+\\
\vspace{0.35 cm}
e^{-\nu}(Q_{8}-4S_{4})(-2Q_{6}+Q_{8}-4S_{4})].\\
\vspace{0.35 cm}
\end{array}
\end{displaymath}

In this case Bianchi identities are reduced to 6 following relations: 
\begin{displaymath}
\begin{array}{c}
\vspace{0.35 cm}
-4r\tilde{F}_{13}+(2r+Q_{9}-4S_{2})(\tilde{F}_{7}-\tilde{F}_{11})+
2\tilde{F}_{11}Q_{7}+\\
\vspace{0.35 cm}
e^{\frac{1}{2}(\lambda-\nu)}[(Q_{8}-4S_{4})(\tilde{F}_{3}-\tilde{F}_{9})+
2\tilde{F}_{9}Q_{6}]-2r^{2}\tilde{F}_{13}'=0,\\
\vspace{0.35 cm}
\end{array}
\end{displaymath}
\begin{equation}
\begin{array}{c}
\vspace{0.35 cm}
e^{\frac{1}{2}(\nu-\lambda)}[(\tilde{F}_{10}-\tilde{F}_{6})(2r+Q_{9}-
4S_{2})-2\tilde{F}_{10}Q_{7}]+\\
\vspace{0.35 cm}
(Q_{8}-4S_{4})(\tilde{F}_{8}-\tilde{F}_{2})-
2F_{8}Q_{6}+2r^{2}\dot{\tilde{F}_{13}}=0,\\
\vspace{0.35 cm}
e^{\frac{1}{2}\nu}[\tilde{F}_{2}(\frac{1}{r}Q_{9}-2-r\nu')+\frac{1}{r}
\tilde{F}_{1}(2r-2Q_{7}+Q_{9}-4S_{2})-2r\tilde{F'}_{2}]-
e^{\frac{1}{2}\lambda-\nu}r\tilde{F}_{3}Q_{0}+\\
\vspace{0.35 cm}
re^{-\frac{1}{2}\nu}
[\tilde{F}_{2}Q_{1}+\tilde{F}_{7}(-Q_{1}-4S_{1}+e^{\nu}\nu')]+
e^{\frac{1}{2}\lambda}\{\frac{1}{r}[-\tilde{F}_{3}Q_{8}+
(\tilde{F}_{12}-\tilde{F}_{0})(2Q_{6}-\\
\vspace{0.35 cm}
Q_{8}+4S_{4})]+
r(2\dot{\tilde{F}_{3}}+\tilde{F}_{3}\dot{\lambda})\}-
e^{-\frac{1}{2}\lambda}r\tilde{F}_{6}(-2Q_{3}+Q_{4}-4S_{3}+e^{\lambda}
\dot{\lambda})=0,\\
\vspace{0.35 cm}
e^{\frac{1}{2}\nu}\{\tilde{F}_{6}(-2-r\nu')-2r\tilde{F'}_{6}+
\frac{1}{r}[\tilde{F}_{6}Q_{9}+(\tilde{F}_{12}+\tilde{F}_{5})
(2r-2Q_{7}+Q_{9}-4S_{2})]\}+\\
\vspace{0.35 cm}
e^{-\frac{1}{2}\nu}r\tilde{F}_{3}(-Q_{1}+2Q_{2}-4S_{1}+e^{\nu}\nu')+
e^{-\frac{1}{2}\lambda}r[\tilde{F}_{7}Q_{4}-\tilde{F}_{2}(Q_{4}-4S_{3}+
e^{\lambda}\dot{\lambda})]-\\
\vspace{0.35 cm}
e^{-\lambda+\frac{1}{2}\nu}r\tilde{F}_{6}Q_{5}+e^{\frac{1}{2}\lambda}
\{\frac{1}{r}[-\tilde{F}_{7}Q_{8}-\tilde{F}_{4}(2Q_{6}-Q_{8}+4S_{4})]+
2r\dot{\tilde{F}_{7}}+r\tilde{F}_{7}\dot{\lambda}\}=0,\\
\vspace{0.35 cm}
e^{\frac{1}{2}\nu}\{2\tilde{F}_{8}+\frac{1}{r}[\tilde{F}_{8}Q_{9}-
\tilde{F}_{4}(2r+Q_{9}-4S_{2})]+2r\tilde{F'}_{8}+r\nu'(\tilde{F}_{8}-
\tilde{F}_{11})\}+\\
\vspace{0.35 cm}
e^{-\frac{1}{2}\nu}r[\tilde{F}_{8}Q_{1}+\tilde{F}_{11}(Q_{1}-2Q_{2}+
4S_{1})]-e^{\frac{1}{2}\lambda-\nu}r\tilde{F}_{9}Q_{0}+
e^{\frac{1}{2}\lambda}\{\frac{1}{r}[(\tilde{F}_{12}-\\
\vspace{0.35 cm}
\tilde{F}_{0})(Q_{8}-4S_{4})-\tilde{F}_{9}Q_{8}]-2r\dot{\tilde{F}_{9}}+
r\dot{\lambda}(\tilde{F}_{10}-\tilde{F}_{9})\}+
e^{-\frac{1}{2}\lambda}r\tilde{F}_{10}(Q_{4}-4S_{3})=0,\\
\vspace{0.35 cm}
e^{\frac{1}{2}\nu}\{\frac{1}{r}[(\tilde{F}_{12}+\tilde{F}_{5})(2r+Q_{9}-
4S_{2})-\tilde{F}_{10}Q_{9}]-2\tilde{F}_{10}-2r\tilde{F'}_{10}+
r\nu'(\tilde{F}_{9}-\tilde{F}_{10})\}-\\
\vspace{0.35 cm}
e^{-\frac{1}{2}\nu}r\tilde{F}_{9}(Q_{1}+4S_{1})+
e^{-\lambda+\frac{1}{2}\nu}r\tilde{F}_{10}Q_{5}+e^{\frac{1}{2}\lambda}
\{\frac{1}{r}[\tilde{F}_{11}Q_{8}+\tilde{F}_{1}(Q_{8}-4S_{4})]+\\
\vspace{0.35 cm}
2r\dot{\tilde{F}_{11}}+r\dot{\lambda}(\tilde{F}_{11}-F_{8})\}-
e^{-\frac{1}{2}\lambda}r[\tilde{F}_{11}Q_{4}+\tilde{F}_{8}(Q_{4}-
2Q_{3}-4S_{3})]=0.\\
\vspace{0.35 cm}
\end{array}
\end{equation}
 
Now let us consider spherically-symmetric gravitational fields in the frame 
of MAGT corresponding to particular gravitational Lagrangian:
\begin{equation}
L_{G}=f_{0}F+fF^{2}+aS_{\nu\mu\alpha}S^{\alpha\mu\nu}+
kQ_{\mu\nu\lambda}Q^{\mu\lambda\nu}+
mQ^{\alpha}{}_{\lambda\alpha}S_{\beta}{}^{\beta\lambda},
\end{equation}
where $f_{0}=(16 \pi G)^{-1}$, G is Neuton's gravitational constant; 
$f, a, k, m$ are indefinite parameters, $F=F^{\mu\nu}{}_{\mu\nu}$.

 The gravitational equations can be obtained by variation of total action
integral
\begin{equation} 
I = \int\delta^{4}xh(L_{G}+L_{m}) 
\end{equation}
($L_{m}$ is Lagrangian of matter and $h = det(h^{i}{}_{\mu})$)
 with respect t® $h^{i}{}_{\mu}$ and 
$A^{ik}{}_{\mu}$. As result of variation we get 16 h-equations:
\begin{equation}
H_{i}{}^{\mu}-\nabla_{\nu}\sigma_{i}{}^{\mu\nu}=t_{i}{}^{\mu},
\end{equation}
and 64 A-equations:
\begin{equation}
2\nabla_{\nu}\varphi_{ik}{}^{\nu\mu}+\sigma_{ik}{}^{\mu}=-J_{ik}{}^{\mu},
\end{equation}
were $H_{i}{}^{\mu}=h^{-1}(\delta L_{G}/\delta h^{i}{}_{\mu})$,   
 $\sigma_{i}{}^{\mu\nu}=(\partial L_{G}/\partial S^{i}{}_{\mu\nu})$,
 $\varphi_{ik}{}^{\mu\nu}=(\partial L_{G}/\partial F^{ik}{}_{\mu\nu})$,
 $t_{i}{}^{\mu}=-h^{-1}(\delta L_{m}/\delta h^{i}{}_{\mu})$,
 $J_{ik}{}^{\mu}=-h^{-1}(\delta L_{m}/\delta A^{ik}{}_{\mu})$.

In spherically-symmetric case with vanishing
 pseudoscalar torsion and nonmetricity functions the system of 
 gravitational equations $(18)-(19)$ is reduced to 19 differential equations:

\begin{displaymath}
\begin{array}{c}
\vspace{0.35 cm}
2f_{0}(\tilde{F}_{19}-\tilde{F}_{20}-\tilde{F}_{17})+f[4\tilde{F}_{19}^{2}-
\tilde{F}_{14}^{2}-(2\tilde{F}_{1}-\tilde{F}_{4})^{2}
+4(\tilde{F}_{20}+\tilde{F}_{17})^{2}-\\
\vspace{0.35 cm}
8\tilde{F}_{19}(\tilde{F}_{20}+
\tilde{F}_{17})+
4\tilde{F}_{8}(2\tilde{F}_{1}-\tilde{F}_{4})-4\tilde{F}_{8}^{2}+
2\tilde{F}_{14}(\tilde{F}_{4}-2\tilde{F}_{1}+2\tilde{F}_{8})]+\\
\vspace{0.35 cm}
k\{e^{-\lambda-2\nu}Q_{2}^{2}-e^{-3\nu}Q_{0}^{2}+e^{-2\lambda-\nu}Q_{3}^{2}-
e^{-3\lambda}Q_{5}^{2}+\frac{1}{r^{4}}[2e^{-\nu}Q_{6}^{2}-\\
\vspace{0.35 cm}
2e^{-\lambda}Q_{7}(Q_{7}+2Q_{9})]\}+
\frac{m}{4}\{e^{-\lambda-2\nu}(Q_{1}Q_{2}+
4Q_{2}S_{1}-Q_{0}Q_{4})+\\
\vspace{0.35 cm}
\frac{4}{r}(e{-\lambda-\nu}Q_{2}-e^{-2\lambda}Q_{5})+
e^{-2\lambda-\nu}(Q_{3}Q_{4}-Q_{1}Q_{5}-4Q_{3}S_{3})+\\
\vspace{0.35 cm}
\frac{2}{r^{2}}[e^{-\lambda-\nu}(Q_{4}Q_{6}-4Q_{6}S_{3}-
Q_{1}Q_{7}+Q_{3}Q_{8}-4Q_{3}S_{4})-\\
\vspace{0.35 cm}
e^{-2\nu}Q_{0}Q_{8}+
4e^{-2\lambda}Q_{5}S_{2}+
e^{-\nu}Q_{6}\dot{\lambda}-2e^{-\lambda}Q'_{7}+
e^{-\lambda}Q_{7}\lambda']+\\
\vspace{0.35 cm}
\frac{4}{r^{4}}[e^{-\nu}Q_{6}(Q_{8}-4S_{4})+4e^{-\lambda}Q_{7}S_{2}]+
e^{-2\nu}Q_{0}\dot{\lambda}+\\
\vspace{0.35 cm}
e^{-2\lambda}(3Q_{5}\lambda'-2Q_{5}')+
e^{-\lambda-\nu}(Q_{3}\dot{\lambda}+
2Q_{2}'-Q_{2}\lambda'-2Q_{2}\nu')\}+\\
\vspace{0.35 cm}
\frac{a}{2}\{e^{-\lambda-\nu}(\frac{4}{r}S_{1}-S_{3}\dot{\lambda}+2S'_{1}-
S_{1}\lambda'-
2S_{1}\nu')+e^{-\lambda-2\nu}(Q_{1}S_{1}+\\
\vspace{0.35 cm}
2S_{1}^{2})-
\frac{1}{r^{4}}[4e^{-\lambda}S_{2}^{2}+2e^{-\nu}(Q_{8}S_{4}-2S_{4}^{2})]+
e^{-2\lambda-\nu}(2S_{3}^{2}-Q_{4}S_{3})\}=t_{\hat{0}}{}^{\hat{0}},\\
\vspace{0.35 cm}
(\tilde{F}_{5}-\tilde{F}_{18})[2f_{0}+4f(2\tilde{F}_{19}-\tilde{F}_{14}-
2\tilde{F}_{20}-2\tilde{F}_{1}+\tilde{F}_{4}-2\tilde{F}_{17}+\\
\vspace{0.35 cm}
2\tilde{F}_{8})]+2k[e^{-\frac{1}{2}(\lambda+5\nu)}Q_{0}Q_{2}-
e^{-\frac{3}{2}(\lambda+\nu)}Q_{2}(Q_{3}+Q_{4})+\\
\vspace{0.35 cm}
e^{-\frac{1}{2}(5\lambda+
\nu)}Q_{4}Q_{5}+\frac{2}{r^{4}}e^{-\frac{1}{2}(\lambda+\nu)}Q_{7}Q_{8}]+
\frac{m}{4}\{e^{-\frac{1}{2}(\lambda+5\nu)}Q_{0}(Q_{2}-\\
\vspace{0.35 cm}
Q_{1}-4S_{1})+
e^{-\frac{3}{2}(\lambda+\nu)}(Q_{3}Q_{1}-2Q_{3}Q_{2}+
Q_{0}Q_{5}+4S_{1}Q_{3}-\\
\vspace{0.35 cm}
4S_{1}Q_{4}+4Q_{2}S_{3})+
\frac{1}{r^{2}}e^{-\frac{1}{2}(\lambda+3\nu)}[2Q_{6}(Q_{1}-2Q_{2}+4S_{1})+\\
\vspace{0.35 cm}
2Q_{0}Q_{7}+2Q_{2}Q_{8}+8Q_{2}S_{4}+
r^{2}(Q_{2}\dot{\lambda}-2\dot{Q}_{2}+
2Q_{2}\dot{\nu}+Q_{0}\nu')]+\\
\vspace{0.35 cm}
e^{-\frac{1}{2}(3\lambda+\nu)}[2\dot{Q}_{5}-
3Q_{5}\dot{\lambda}-
\frac{1}{r^{2}}(2Q_{5}Q_{8}+Q_{4}S_{2})-Q_{3}\nu']+\\
\vspace{0.35 cm}
\frac{1}{r^{2}}e^{-\frac{1}{2}(\lambda+\nu)}(4\dot{Q}_{7}-
\frac{4}{r^{2}}Q_{7}Q_{8}-2Q_{7}\dot{\lambda}-Q_{6}\nu')\}+\\
\vspace{0.35 cm}
\frac{a}{2}[\frac{2}{r^{4}}e^{-\frac{1}{2}(\lambda+\nu)}Q_{8}S_{2}-
e^{-\frac{1}{2}(\lambda+5\nu)}Q_{0}S_{1}+
e^{-\frac{3}{2}(\lambda+\nu)}S_{3}(2Q_{2}-Q_{1}-4S_{1})+\\
\vspace{0.35 cm}
e^{-\frac{1}{2}(\lambda+3\nu)}(S_{1}\dot{\lambda}-2\dot{S}_{1}+
2S_{1}\dot{\nu})+e^{-\frac{1}{2}(3\lambda+\nu)}S_{3}\nu']=
t_{\hat{1}}{}^{\hat{0}},\\
\vspace{0.35 cm}
\end{array}
\end{displaymath}
\begin{displaymath}
\begin{array}{c}
\vspace{0.35 cm}
2(\tilde{F}_{15}-\tilde{F}_{9})[f_{0}+2f(2\tilde{F}_{19}-\tilde{F}_{14}-
2\tilde{F}_{20}-2\tilde{F}_{1}+2\tilde{F}_{4}-2\tilde{F}_{17}+\\
\vspace{0.35 cm}
2\tilde{F}_{8})]+2k[e^{-\frac{3}{2}(\lambda+\nu)}Q_{3}(O_{1}+
Q_{2})-e^{-\frac{1}{2}(\lambda+5\nu)}Q_{0}Q_{1}-\\
\vspace{0.35 cm}
e^{-\frac{1}{2}(5\lambda+\nu)}Q_{3}Q_{5}-
\frac{2}{r^{4}}e^{-\frac{1}{2}(\lambda+\nu)}Q_{6}Q_{9}]+
\frac{m}{4}\{e^{-\frac{3}{2}(\lambda+\nu)}(2Q_{2}Q_{3}-\\
\vspace{0.35 cm}
Q_{2}Q_{4}-Q_{0}Q_{5}+4Q_{3}S_{1}-4Q_{1}S_{3}+4Q_{2}S_{3})+
e^{-\frac{1}{2}(5\lambda+\nu)}Q_{5}(Q_{4}-Q_{3}-4S_{3})+\\
\vspace{0.35 cm}
e^{-\frac{1}{2}(\lambda+\nu)}[\frac{8}{r^{3}}Q_{6}+
\frac{4}{r^{4}}Q_{6}Q_{9}+
\frac{2}{r^{2}}(Q_{7}\dot{\lambda}-2Q'_{6}+Q_{6}\nu')]+
e^{-\frac{1}{2}(3\lambda+\nu)}[\frac{2}{r^{2}}(Q_{5}Q_{6}-\\
\vspace{0.35 cm}
2Q_{3}Q_{7}+Q_{4}Q_{7}+Q_{3}Q_{9}+4Q_{3}S_{2}-
4Q_{7}S_{3})+Q_{5}\dot{\lambda}-2Q'_{3}+2Q_{3}\lambda'+Q_{3}\nu']+\\
\vspace{0.35 cm}
e^{-\frac{1}{2}(\lambda+3\nu)}[2Q'_{0}-3Q_{0}\nu'-
\frac{2}{r^{2}}(Q_{0}Q_{9}+4Q_{1}S_{4})-Q_{2}\dot{\lambda}]\}+
\frac{a}{2}[e^{-\frac{3}{2}(\lambda+\nu)}S_{1}(2Q_{3}-\\
\vspace{0.35 cm}
Q_{4}+4S_{3})+e^{-\frac{1}{2}(3\lambda+\nu)}(\frac{4}{r}S_{3}+2S'_{3}+
2S_{3}\lambda'-S_{3}\nu')+e^{-\frac{1}{2}(\lambda+\nu)}Q_{5}S_{3}-\\
\vspace{0.35 cm}
e^{-\frac{1}{2}(\lambda+\nu)}(\frac{4}{r^{3}}S_{4}+
\frac{2}{r^{4}}Q_{9}S_{4})-e^{-\frac{1}{2}(\lambda+3\nu)}S_{1}\dot{\lambda}]=
t_{\hat{0}}{}^{\hat{1}},\\
\vspace{0.35 cm}
2f_{0}(\tilde{F}_{8}-\tilde{F}_{1}-\tilde{F}_{20})+
f[8\tilde{F}_{20}\tilde{F}_{1}-\tilde{F}_{14}^{2}-4\tilde{F}_{19}^{2}+
4\tilde{F}_{1}^{2}-4\tilde{F}_{19}(\tilde{F}_{4}-2\tilde{F}_{17})+\\
\vspace{0.35 cm}
2\tilde{F}_{14}(2\tilde{F}_{19}+\tilde{F}_{4}-2\tilde{F}_{7})-
(\tilde{F}_{4}-2\tilde{F}_{17})^{2}-8\tilde{F}_{1}\tilde{F}_{8}+
4(\tilde{F}_{20}-\tilde{F}_{8})^{2}]+\\
\vspace{0.35 cm}
k[e^{-3\nu}Q_{0}^{2}-e^{-\lambda-2\nu}Q_{2}^{2}-e^{-2\lambda-\nu}Q_{3}^{2}+
e^{-3\lambda}Q_{5}^{2}+\frac{1}{r^{4}}e^{-\nu}(2Q_{6}^{2}+4Q_{6}Q_{8})-\\
\vspace{0.35 cm}
\frac{2}{r^{4}}e^{-\lambda}Q_{7}^{2}]+
m\{e^{-\lambda-\nu}[\frac{1}{r}Q_{2}-\frac{1}{2r^{2}}(Q_{4}Q_{6}-Q_{1}Q_{7}-
Q_{2}Q_{9}-4Q_{7}S_{1}+\\
\vspace{0.35 cm}
4Q_{2}S_{2})+\frac{1}{2}\dot{Q}_{3}-
\frac{1}{2}Q_{3}\dot{\lambda}-
\frac{1}{4}Q_{3}\dot{\nu}+\frac{1}{4}Q_{2}\nu']+\\
\vspace{0.35 cm}
\frac{1}{4}e^{-\lambda-2\nu}(Q_{0}Q_{4}-Q_{1}Q_{2}-4Q_{2}S_{1})+
\frac{1}{4}e^{-2\lambda-\nu}(Q_{1}Q_{5}-Q_{3}Q_{4}+\\
\vspace{0.35 cm}
4Q_{3}S_{3})-
e^{-2\lambda}Q_{5}(\frac{1}{r}+\frac{1}{2r^{2}}Q_{9}+\frac{1}{4}\nu')+
e^{-\lambda}Q_{7}(\frac{4}{r^{4}}S_{2}-\frac{2}{r^{3}}-
\frac{1}{r^{4}}Q_{9}-\frac{1}{2r^{2}}\nu')+\\
\vspace{0.35 cm}
e^{-2\nu}(\frac{2}{r^{2}}Q_{0}S_{4}-\frac{1}{2}\dot{Q}_{0}+
\frac{3}{4}Q_{0}\dot{\nu})+\frac{1}{r^{2}}e^{-\nu}(\dot{Q}_{6}-
\frac{4}{r^{2}}Q_{6}S_{4}-\frac{1}{2}Q_{6}\dot{\nu})\}+
a[\frac{1}{r^{3}}e^{-\lambda}S_{2}(2+\\
\vspace{0.35 cm}
\frac{1}{r}Q_{9}-\frac{2}{r}S_{2})-e^{-\lambda-2\nu}(\frac{1}{2}Q_{1}S_{1}+
S_{1}^{2})+e^{-2\lambda-\nu}(\frac{1}{2}Q_{4}S_{3}-S_{3}^{2})+
\frac{2}{r^{4}}e^{-\nu}S_{4}^{2}+\\
\vspace{0.35 cm}
e^{-\lambda-\nu}(S_{3}\dot{\lambda}-\dot{S}_{3}+\frac{1}{2}S_{3}\dot{\nu}+
\frac{1}{2}S_{1}\nu')]=t_{\hat{1}}{}^{\hat{1}},\\
\vspace{0.35 cm}
\end{array}
\end{displaymath}
\begin{displaymath}
\begin{array}{c}
\vspace{0.35 cm}
f_{0}(\tilde{F}_{4}-\tilde{F}_{1}+\tilde{F}_{19}-\tilde{F}_{17}+
\tilde{F}_{8}-\tilde{F}_{14})+f[\tilde{F}_{14}^{2}-4\tilde{F}_{20}^{2}-\\
\vspace{0.35 cm}
4\tilde{F}_{20}\tilde{F}_{1}-2\tilde{F}_{1}\tilde{F}_{4}+\tilde{F}_{4}^{2}+
(4\tilde{F}_{20}+2\tilde{F}_{4})(\tilde{F}_{19}-\tilde{F}_{17}+
\tilde{F}_{8})-\\
\vspace{0.35 cm}
2\tilde{F}_{14}(\tilde{F}_{19}-\tilde{F}_{1}+\tilde{F}_{4}-
\tilde{F}_{17}+\tilde{F}_{8})]+k[e^{-3\nu}Q_{0}^{2}-\\
\vspace{0.35 cm}
e^{-\lambda-2\nu}(2Q_{1}Q_{2}+Q_{2}^{2})+e^{-2\lambda-\nu}(Q_{3}^{2}+
2Q_{3}Q_{4})-e^{-3\lambda}Q_{5}^{2}+\frac{2}{r^{4}}e^{-\nu}Q_{6}Q_{8}-\\
\vspace{0.35 cm}
\frac{2}{r^{4}}e^{-\lambda}Q_{7}Q_{9}]+
m\{e^{-\lambda-\nu}[\frac{1}{2r}Q_{2}+\frac{1}{r^{2}}(Q_{7}S_{1}-
\frac{1}{4}Q_{3}Q_{8}-\frac{1}{4}Q_{2}Q_{9}-\\
\vspace{0.35 cm}
Q_{2}S_{2}-Q_{6}S_{3}-
Q_{3}S_{4})+\frac{1}{2}\dot{Q}_{3}-
\frac{1}{4}Q_{3}\dot{\lambda}-\frac{1}{4}Q_{3}\dot{\nu}+
\frac{1}{2}Q'_{2}-\\
\vspace{0.35 cm}
\frac{1}{4}Q_{2}\lambda'-\frac{1}{4}Q_{2}\nu']+
e^{-2\lambda}[Q_{5}(\frac{1}{4r^{2}}Q_{9}+\frac{1}{r^{2}}S_{2}-
\frac{1}{2r}+\\
\vspace{0.35 cm}
\frac{3}{4}\lambda'-\frac{1}{4}\nu')-\frac{1}{2}Q'_{5}]+
e^{-\lambda}[Q_{7}(\frac{1}{r^{3}}+\frac{1}{2r^{4}}Q_{9}+\\
\vspace{0.35 cm}
\frac{1}{2r^{2}}\lambda'-
\frac{1}{2r^{2}}\nu')-\frac{1}{r^{2}}Q'_{7}]+
e^{-2\nu}[Q_{0}(\frac{1}{4r^{2}}Q_{8}+
\frac{1}{r^{2}}S_{4}-
\frac{1}{4}\dot{\lambda}+\frac{3}{4}\dot{\nu})-\\
\vspace{0.35 cm}
\frac{1}{2}\dot{Q}_{0}]+
\frac{1}{2r^{2}}e^{-\nu}[2\dot{Q}_{6}+Q_{6}(\dot{\lambda}-\dot{\nu}-
\frac{1}{r^{2}}Q_{8})]+e^{-\lambda-2\nu}(Q_{0}S_{3}-Q_{2}S_{1})+\\
\vspace{0.35 cm}
e^{-2\lambda-\nu}(Q_{5}S_{1}-Q_{3}S_{3})\}+
a[\frac{1}{2r^{2}}e^{-\lambda}(2S'_{2}-S_{2}\lambda'+
S_{2}\nu'-\frac{2}{r}S_{2}-\frac{1}{r^{2}}Q_{9}S_{2})-\\
\vspace{0.35 cm}
e^{-\lambda-2\nu}S_{1}^{2}+e^{-2\lambda-\nu}S_{3}^{2}+
\frac{1}{2r^{2}}e^{-\nu}(\frac{1}{r^{2}}Q_{8}S_{4}-2\dot{S}_{4}-
S_{4}\dot{\lambda}+S_{4}\dot{\nu})]=t_{\hat{2}}{}^{\hat{2}},\\
\vspace{0.35 cm}
4e^{-\frac{3}{2}\nu}kQ_{0}+(e^{-\lambda-\frac{1}{2}\nu}Q_{3}+
\frac{2}{r^{2}}e^{-\frac{1}{2}\nu}Q_{6})[f_{0}+2f(2F_{19}-F_{14}-2F_{20}-\\
\vspace{0.35 cm}
2F_{1}+F_{4}-2F_{17}+2F_{8})]=J_{00}{}^{0},\\
\vspace{0.35 cm}
-\frac{1}{2}e^{-\frac{1}{2}\lambda-\nu}Q_{2}[2f_{0}+4f(2F_{19}-F_{14}-
2F_{20}-2F_{1}+F_{4}-2F_{17}+2F_{8})-\\
\vspace{0.35 cm}
1+8k]-\frac{1}{2}e^{-\frac{3}{2}\lambda}Q_{5}-
\frac{1}{r^{2}}e^{-\frac{1}{2}\lambda-\nu}(e^{\nu}Q_{7}-ar^{2}S_{1})=
J_{00}{}^{1},\\
\vspace{0.35 cm}
\frac{1}{2}e^{-\frac{1}{2}\lambda-\nu}[4k(Q_{1}+Q_{2})+2aS_{1}]+
[\frac{1}{2}e^{-\frac{1}{2}\lambda-\nu}Q_{1}+
\frac{1}{2}e^{-\frac{3}{2}\lambda}Q_{5}+
\frac{1}{r^{2}}e^{-\frac{1}{2}\lambda}(2Q_{7}-Q_{9}+\\
\vspace{0.35 cm}
4S_{2})][f_{0}+2f(2F_{19}-F_{14}-2F_{20}-2F_{1}+F_{4}-2F_{17}+2F_{8})]+
2fe^{-\frac{1}{2}\lambda}(2F'_{19}-\\
F'_{14}-2F'_{20}-2F'_{1}+F'_{4}-2F'_{17}+2F'_{8})=J_{01}{}^{0},\\
\vspace{0.35 cm}
[\frac{1}{r^{2}}e^{-\frac{1}{2}\nu}(Q_{8}-2S_{4})-
\frac{1}{2}e^{-\frac{3}{2}\nu}Q_{0}-
\frac{1}{2}e^{-\lambda-\frac{1}{2}\nu}Q_{4}]
[f_{0}+2f(2F_{19}-F_{14}-2F_{20}-\\
\vspace{0.35 cm}
2F_{1}+F_{4}-2F_{17}+2F_{8})]-2e^{-\lambda-\frac{1}{2}\nu}k(Q_{3}+Q_{4})+
\frac{1}{2}e^{-\lambda-\frac{1}{2}\nu}S_{3}+
\frac{1}{r^{2}}e^{-\frac{1}{2}\nu}S_{4}-\\
\vspace{0.35 cm}
fe^{-\frac{1}{2}\nu}(2\dot{F}_{19}-\dot{F}_{14}-
2\dot{F}_{20}-2\dot{F}_{1}+\dot{F}_{4}-2\dot{F}_{17}+2\dot{F}_{8})=
J_{01}{}^{1},\\
\vspace{0.35 cm}
(\frac{1}{2}e^{-\lambda-\frac{1}{2}\nu}Q_{4}-
\frac{1}{2}e^{-\frac{3}{2}\nu}Q_{0}-
2e^{-\lambda-\frac{1}{2}\nu}S_{3}-
\frac{2}{r^{2}}e^{-\frac{1}{2}\nu}S_{4})[f_{0}+2f(2F_{19}-F_{14}-2F_{20}-\\
\vspace{0.35 cm}
2F_{1}+F_{4}-2F_{17}+2F_{8})]-
\frac{2}{r^{2}}e^{-\frac{1}{2}\nu}k(Q_{6}+Q_{8})+
\frac{4}{r^{2}}e^{-\frac{1}{2}\nu}S_{4}+
2e^{-\lambda-\frac{1}{2}\nu}S_{3}-\\
\vspace{0.35 cm}
2fe^{-\frac{1}{2}\nu}(2\dot{F}_{19}-
\dot{F}_{14}-2\dot{F}_{20}-2\dot{F}_{1}+
\dot{F}_{4}-2\dot{F}_{17}+2\dot{F}_{8})=J_{02}{}^{2},\\
\vspace{0.35 cm}
\end{array}
\end{displaymath}
\begin{equation}
\begin{array}{c}
\vspace{0.35 cm}
(\frac{1}{2}e^{-\frac{1}{2}\lambda-\nu}Q_{1}+
\frac{1}{2}e^{-\frac{3}{2}\lambda}Q_{5}+
\frac{1}{r^{2}}e^{-\frac{1}{2}\lambda}Q_{9}-
\frac{4}{r^{2}}e^{-\frac{1}{2}\lambda}S_{2})[f_{0}+
2f(2F_{19}-F_{14}-2F_{20}-\\
\vspace{0.35 cm}
2F_{1}+F_{4}-2F_{17}+2F_{8})]+
2e^{-\frac{1}{2}\lambda-\nu}k(Q_{1}+Q_{2})-
2fe^{-\frac{1}{2}\lambda}(2F'_{19}-F'_{14}-\\
\vspace{0.35 cm}
2F'_{20}-2F'_{1}+F'_{4}-2F'_{17}+2F'_{8})=J_{10}{}^{0},\\
\vspace{0.35 cm}
[\frac{1}{r^{2}}e^{-\frac{1}{2}\nu}(2Q_{6}-Q_{8}+4S_{4})-
\frac{1}{2}e^{-\frac{3}{2}\nu}Q_{0}-\frac{1}{2}e^{-\lambda-
\frac{1}{2}\nu}Q_{4}][f_{0}+\\
\vspace{0.35 cm}
2f(2F_{19}-F_{14}-2F_{20}-
2F_{1}+F_{4}-2F_{17}+2F_{8})]+\\
\vspace{0.35 cm}
2fe^{-\frac{1}{2}\nu}(2\dot{F}_{19}-
\dot{F}_{14}-2\dot{F}_{20}-2\dot{F}_{1}+
\dot{F}_{4}-2\dot{F}_{17}+2\dot{F}_{8})+
\frac{1}{2}e^{-\frac{3}{2}\nu}Q_{0}-\\
\vspace{0.35 cm}
\frac{1}{2}e^{-\lambda-\frac{1}{2}\nu}[Q_{3}(1+4k)+4kQ_{4}-2S_{3}(1+a)]+
\frac{1}{r^{2}}e^{-\frac{1}{2}\nu}(2S_{4}-Q_{6})=J_{10}{}^{1},\\
\vspace{0.35 cm}
e^{-\lambda-\frac{1}{2}\nu}Q_{3}[f_{0}+2f(2F_{19}-F_{14}-2F_{20}-
2F_{1}+F_{4}-2F_{17}+2F_{8})]+\\
\vspace{0.35 cm}
e^{-\lambda-\frac{1}{2}\nu}(4kQ_{3}+aS_{3})=J_{11}{}^{0},\\
\vspace{0.35 cm}
(\frac{2}{r^{2}}e^{-\frac{1}{2}\lambda}Q_{7}-
e^{-\frac{1}{2}\lambda-\nu}Q_{2})[f_{0}+2f(2F_{19}-F_{14}-2F_{20}-
2F_{1}+F_{4}-\\
\vspace{0.35 cm}
2F_{17}+2F_{8})]-
4e^{-\frac{3}{2}\lambda}kQ_{5}-2e^{-\frac{1}{2}\lambda-\nu}S_{1}-
\frac{4}{r^{2}}e^{-\frac{1}{2}\lambda}S_{2}=J_{11}{}^{1},\\
\vspace{0.35 cm}
[\frac{1}{2}e^{-\frac{3}{2}\lambda}Q_{5}-
\frac{1}{2}e^{-\frac{1}{2}\lambda-\nu}(Q_{1}+4S_{1})-
\frac{2}{r^{2}}e^{-\frac{1}{2}\lambda}S_{2}][f_{0}+2f(2F_{19}-F_{14}-\\
\vspace{0.35 cm}
2F_{20}-2F_{1}+F_{4}-2F_{17}+2F_{8})]-
2fe^{-\frac{1}{2}\lambda}(2F'_{19}-F'_{14}-2F'_{20}-2F'_{1}+\\
\vspace{0.35 cm}
F'_{4}-2F'_{17}+2F'_{8})+
\frac{2}{r^{2}}e^{-\frac{1}{2}\lambda}[2S_{2}-k(Q_{7}+Q_{9})]+
2e^{-\frac{1}{2}\lambda-\nu}S_{1}=J_{12}{}^{2},\\
\vspace{0.35 cm}
[\frac{1}{2}e^{-\lambda-\frac{1}{2}\nu}(2Q_{3}-Q_{4}+4S_{3})-
\frac{1}{2}e^{-\frac{3}{2}\nu}Q_{0}+
\frac{1}{r^{2}}e^{-\frac{1}{2}\nu}(Q_{6}-Q_{8}+2S_{4})][f_{0}+\\
\vspace{0.35 cm}
2f(2F_{19}-
F_{14}-2F_{20}-2F_{1}+F_{4}-2F_{17}+2F_{8})]+
2fe^{-\frac{1}{2}\nu}(2\dot{F}_{19}-\\
\vspace{0.35 cm}
\dot{F}_{14}-2\dot{F}_{20}-
2\dot{F}_{1}+\dot{F}_{4}-
2\dot{F}_{17}+2\dot{F}_{8})+e^{-\frac{3}{2}\nu}Q_{0}+
e^{-\lambda-\frac{1}{2}\nu}(2S_{3}-Q_{3})+\\
\vspace{0.35 cm}
\frac{1}{r^{2}}e^{-\frac{1}{2}\nu}[2Q_{8}-
2Q_{6}(1+k)+S_{4}(1+a)]=J_{20}{}^{2},\\
\vspace{0.35 cm}
[f_{0}+2f(2F_{19}-F_{14}-2F_{20}-2F_{1}+F_{4}-2F_{17}+
2F_{8})]\\
\vspace{0.35 cm}
[e^{-\frac{1}{2}\lambda-\nu}(\frac{1}{2}Q_{1}-Q_{2}+2S_{1})+
\frac{1}{2}e^{-\frac{3}{2}\lambda}Q_{5}+
\frac{1}{r^{2}}e^{-\frac{1}{2}\lambda}(Q_{7}-\\
\vspace{0.35 cm}
Q_{9}+2S_{2})]+
2fe^{-\frac{1}{2}\lambda}(2F'_{19}-F'_{14}-2F'_{20}-2F'_{1}+F'_{4}-\\
\vspace{0.35 cm}
2F'_{17}+2F'_{8})+e^{-\frac{1}{2}\lambda-\nu}(Q_{2}+2S_{1})-
e^{-\frac{3}{2}\lambda}Q_{5}-\\
\vspace{0.35 cm}
\frac{1}{r^{2}}e^{-\frac{1}{2}\lambda}[2Q_{7}(1+k)+2kQ_{9}-S_{2}(4+a)]=
J_{21}{}^{2},\\
\vspace{0.35 cm}
\frac{1}{r^{2}}e^{-\frac{1}{2}\nu}(4kQ_{6}+aS_{4})+
\frac{1}{r^{2}}e^{-\frac{1}{2}\nu}Q_{6}[f_{0}+2f(2F_{19}-\\
\vspace{0.35 cm}
F_{14}-2F_{20}-2F_{1}+F_{4}-2F_{17}+2F_{8})]=J_{22}{}^{0},\\
\vspace{0.35 cm}
\end{array}
\end{equation}
\begin{displaymath}
\begin{array}{c}
\vspace{0.35 cm}
\frac{1}{2}e^{-\frac{3}{2}\lambda}Q_{5}-
\frac{1}{2}e^{-\frac{1}{2}\lambda-\nu}Q_{2}-
\frac{1}{r^{2}}e^{-\frac{1}{2}\lambda}[Q_{7}(4k-1)+aS_{2}]-\\
\vspace{0.35 cm}
\frac{1}{r^{2}}e^{-\frac{1}{2}\lambda}Q_{7}[f_{0}+2f(2F_{19}-
F_{14}-2F_{20}-2F_{1}+F_{4}-2F_{17}+2F_{8})]=J_{22}{}^{1}.\\
\vspace{0.35 cm}
\end{array}
\end{displaymath}
In the case $t_{i}{}^{\mu}=0$, $J_{ik}{}^{\mu}=0$ 
this system of equations is satisfied by vanishing torsion and nonmetricity  
and vacuum Schwarzchild metrics:
\begin{equation}
g_{\mu\nu}={\rm diag}((1-r_{g}/r), -(1-r_{g}/r)^{-1}, 
-r^{2}, -r^{2}\sin^{2}\theta),  (r_{g}=const).
\end{equation}

\end{document}